\documentclass{ws-ijmpa}
\usepackage[super,compress]{cite}
\usepackage{graphicx}
\usepackage{amsmath}
\usepackage{mathrsfs}
\usepackage{multirow}
\usepackage[figuresright]{rotating}
\usepackage{arydshln}
\usepackage{color}

\newcommand{\pT}{$p_{\mathrm{T}}$} 
\newcommand{\ET}{$E_{\mathrm{T}}$} 
\newcommand{\ee}{$ee$}%
\newcommand{\ep}{$ep$}%
\newcommand{\pp}{$pp$}%
\newcommand{\ppb}{$p\bar{p}$}%
\newcommand{\gp}{$\gamma p$}%
\newcommand{\gamg}{$\gamma \gamma$}%

\newcommand{\nlojetpp}{NLOJET++}%
\newcommand{\alphas}{$\alpha_{\mathrm{S}}$}%
\newcommand{\pythia}{Pythia}%
\newcommand{\herwigpp}{Herwig++}%
\newcommand{\powheg}{Powheg}%

\begin{document}
\markboth{Paolo Francavilla}
{Measurements of inclusive jet and dijet cross sections at the Large Hadron Collider}

%
\catchline{}{}{}{}{}
%

\title{Measurements of inclusive jet and dijet cross sections at the Large Hadron Collider\footnote{Contribution to ``Jet Measurements at the LHC", G. Dissertori ed. To appear in International
Journal of Modern Physics A (IJMPA).}}

\author{Paolo Francavilla}

\address{Laboratoire de Physique Nucl\'eaire et de Hautes Energies and Institute Lagrange de Paris, Universit\'e Pierre et Marie Curie , 7, quai Saint Bernard \\
Paris, France\\
paolo.francavilla@cern.ch}



\maketitle


\begin{abstract}
This review discusses the measurements of the inclusive jet and dijet cross section performed by the experimental collaborations at the LHC during what is now being called LHC Run 1 (2009 -- 2013). It summarises some of the experimental challenges and the techniques used in the measurements of jets cross sections during the LHC Run 1. 

\end{abstract}

\ccode{PACS numbers:}


\section{Introduction}
The measurement of jet production in proton-proton collisions is an important part of the physics program for the experiments at the Large Hadron Collider~\cite{1748-0221-3-08-S08001} (LHC).
Jet production has been measured at 
\ee, \ep, \ppb, and \pp~ colliders, as well as in \gp~ and \gamg~ collisions. These measurements provide a precise determination 
of the strong coupling constant, they are used to obtain information about the structure of the proton and photon, 
and  they are important tools for understanding the strong interaction and searching for physics beyond the 
Standard Model (see, for example Ref. \citen{Arnison:1983dk,Alitti:1990aa,Adeva:1990nu,Chekanov:2001bw,Heister:2002tq,Chekanov:2002be,Abdallah:2004uq,Abbiendi:2005vd,Chekanov:2005nn,Abulencia:2007ez,:2007jx,:2008hua,Aaltonen:2008eq,Abazov:2009nc,:2009he,Aaron:2009vs,:2009mh,Abramowicz:2010ke,Abazov:2010fr}).

The ALICE, ATLAS and CMS Collaborations have measured inclusive jet cross sections at centre-of-mass energies, $\sqrt{s}= 2.76$ TeV \cite{Abelev:2013fn,Aad:2013lpa} and  $\sqrt{s}=7$ TeV \cite{Aad:2010wv, daCosta:2011ni,Aad:2011fc,Aad:2014vwa,Aad:2013tea,Khachatryan:2011zj,CMS:2011ab,Chatrchyan:2011qta,Chatrchyan:2012bja,Chatrchyan:2012gwa,Chatrchyan:2014gia} and performed preliminary measurements at $\sqrt{s}=8$ TeV \cite{CMS:2013kda,CMS:2013tea,CMS:2014aga,CMS:2015aya} during what is now being called LHC Run 1 (2009 -- 2013) . 

Jets in particle physics are  narrow sprays of particles produced in high energy collisions, a footprint of the underlying high energy interactions between quarks and gluons.
The particles in this narrow spray are associated together by a well defined procedure called jet algorithm (see Ref. \citen{Cacciari:2015jwa} for a review).
The collection of particles belonging to a specific jet determines its properties (such as the jet four-momentum) which are used for the measurement of the differential cross sections.

The first elegant measurement of the production of jets consists in counting how many jets have been produced in a given acceptance interval per unity of integrated luminosity. Thus, in this measurement, called the inclusive jet cross section, events containing two jets contribute twice if both jets fall within the specified acceptance region.

As expected from the dynamics of the perturbative regime of the strong interactions  (pQCD), the majority of the events containing jets with sufficiently high transverse momentum (\pT~$>$ 20 GeV) comprises two or more jets. For sufficiently \pT, the two leading jets in \pT~are usually roughly back-to-back in azimuth, and balanced in \pT. 
Dijet cross section measurements are designed to give a description of these topologies. In events with more than one jet, the distributions of the dijet invariant mass, and the differences in the azimuth angle of the system composed by the two leading jets, are measured.

The high rate for the jet production, and the relatively clean and simple definition of the inclusive and dijet jet cross sections made these measurements some of the first high \pT~measurements done at the LHC, and one of the first giving a glimpse into the physics at the TeV scale. 

This review, based on the measurements of the inclusive and dijet  cross sections in proton-proton collisions performed by the LHC Collaborations, is structured as follow:
Section \ref{sec:definition} gives a short review of the measured cross sections and their definition; Section \ref{sec:jetreco} is a review of  the experimental techniques used by the LHC Collaborations to measure jets;
Section \ref{sec:eventsel} describes how jets are measured, while Section \ref{sec:systematics} describes techniques adopted to remove the detector effects from the experimental measurements, and the sources of systematic uncertainties. Section \ref{ThePred} describes some of the improvements on the theoretical tools available to calculate the jet cross section.
Finally Section \ref{sec:measurements} presents some of the measured cross sections, their theoretical uncertainties, and a description of the tests performed on them.

\section{Cross section definition}
\label{sec:definition}
The jet cross sections have been measured by the LHC Collaborations defining jets with the anti-$k_{\mathrm{t}}$ jet algorithm\cite{Cacciari:2008gp}.
The anti-$k_{\mathrm{t}}$ jet algorithm performs a sequential recombination of the constituents of the jets, using their kinematic properties and their relative distance $\Delta R=\sqrt{\Delta \phi^{2} + \Delta y^{2}}$, where $\phi$ is the azimuthal angle of the constituent, and  $y$ is its rapidity. The algorithm uses a recombination parameter $R$ to resolve the jets.

The favourite recombination parameters $R$ used in the measurements of the jet cross sections by the LHC Collaborations are in the range 0.2-0.7, being 0.4 and 0.5 the most widely used. 
It is particularly interesting to measure the inclusive jet and dijet cross sections with different recombination parameters $R$,  because the dynamics of the strong interactions in the perturbative and non perturbative regime (like the effect of the final state radiation, hadronization and underlying event), and the presence of additional proton-proton interactions within the same or neighboring bunch crossings (in-time and out-of-time pile-up) play different roles for different $R$.

The inclusive jet cross section measurement has been performed  as a function of  \pT~and rapidity ($y$) of the jet. 
The double differential inclusive jet cross section is defined as:
\begin{equation}
\frac{d^2\sigma_{incl.~jet}}{d p_{\mathrm{T}} dy} = \frac{N_{jets}}{ \mathscr{L} \Delta p_{\mathrm{T}} \Delta y }
\end{equation}
where $\sigma_{incl.~jet}$ is the inclusive jet cross section, $N_{jets}$ is the number of jets after all corrections (acceptance and efficiency) in a specific bin of rapidity $y$ and \pT~with widths $ \Delta y $ and  $\Delta$\pT~respectively and  $\mathscr{L}$ is the integrated luminosity used in the measurement.

Once the two  leading jets ordered in \pT~are identified, the double differential dijet cross section can be defined as a function of the invariant mass $m_{jj}$ of the dijet system, and an angular variable of the system ($x_{angle}$).
The angular variable has been defined by the LHC Collaborations in different ways: 
\begin{itemize}
\item as the maximum of the rapidities of the two selected jets:  \\ $x_{angle}=y_{max}=\mathrm{max}(|y_{1}|, |y_{2}|)$; 
\item as  half of the rapidity separation of the two selected jets: \\ $x_{angle}=y*=|y_{1} - y_{2} |$;
\item as the variable $\chi_{jj}$ defined as: \\ $x_{angle}=\chi_{jj}=\mathrm{exp}(|y_{1} - y_{2} |)$.
\end{itemize}

The double differential dijet cross section is:
\begin{equation}
\frac{d^2\sigma_{dijet}}{d m_{jj} dx_{angle}} = \frac{N_{dijet}}{ \mathscr{L} \Delta m_{jj} \Delta x_{angle}}
\end{equation}
where $\sigma_{dijet}$ is the dijet cross section,  $N_{dijet}$ is the number of events in which the  dijet system has a $m_{jj}$  and a $x_{angle}$ in a certain bin, with widths $ \Delta m_{jj} $ and  $\Delta x_{angle}$ respectively, and,  $\mathscr{L}$ is the integrated luminosity used in the measurement.

In addition, the dijet azimuthal correlation has been measured as:
\begin{equation}
\frac{1}{\sigma_{dijet}}~\frac{d^2\sigma_{dijet}}{d \phi_{jj}}=\frac{1}{N_{tot}} ~ \frac{N_{dijet}}{ \Delta \phi_{jj}} 
\end{equation}
where $\Delta \phi_{jj}$ is the angular separation in the transverse plane between the two leading jets.
In this case, the measurement has been performed in different ranges of \pT$^{max}$ where  \pT$^{max}$  is the maximum of the \pT~of the two leading jets.

A list of all the inclusive jet and dijet cross section measurements performed by the LHC collaborations in Run1 is shown in Table \ref{table:paper}.
\begin{table}[h!]
\tbl{Inclusive jet cross sections and dijet cross sections measurements performed by the LHC collaborations in Run1.}{
\rotatebox{90}{
\begin{tabular}{l |l |l |l |l |l}
Collab. & Dataset & Measurement & anti-$k_{\mathrm{t}}$  $R$ & Kinematic ranges & Ref.  \\ \hline \hline
ALICE & 2.76 TeV, $\mathscr{L}=$ 13.6 nb$^{-1}$ & inclusive jet &  0.2 and 0.4 & ($p_{\mathrm{T}}/\mathrm{GeV}  \geq 20$) $\otimes~(|\eta| < 0.5$)  &\citen{Abelev:2013fn}  \\ \hline
 & 2.76 TeV, $\mathscr{L}=$ 0.20 pb$^{-1}$ & inclusive jet &  0.4 and 0.6 & ($ p_{\mathrm{T}}/\mathrm{GeV} \geq 20$) $\otimes~(|y| < 4.4$) &  \citen{Aad:2013lpa}\\ \cdashline{2-6}
 \multirow{8}{*}{ATLAS} &  \multirow{2}{*}{7 TeV, $\mathscr{L}=$ 17 nb$^{-1}$}& inclusive jet  & \multirow{2}{*}{0.4 and 0.6} &  ($ p_{\mathrm{T}}/\mathrm{GeV} \geq 60$) $\otimes~(|y| < 2.8$) & \multirow{2}{*}{\citen{Aad:2010wv}}\\ 
 &  & dijet mass &  & for dijet mass: ($ p^{2nd}_{\mathrm{T}}/\mathrm{GeV} \geq 30$) &  \\ \cdashline{2-6}
  & \multirow{2}{*}{7 TeV, $\mathscr{L}=$ 36 pb$^{-1}$} &  \multirow{2}{*}{$\Delta \phi_{jj} $} &   \multirow{2}{*}{0.4} & ($ p_{\mathrm{T}}/\mathrm{GeV} \geq 100$) $\otimes~(|y| < 2.8$)& \multirow{2}{*}{\citen{daCosta:2011ni}} \\ 
 & &  & & ($ p^{max}_{\mathrm{T}}/\mathrm{GeV} \geq 110$) $\otimes~(|y^{1st,2nd}| < 0.8$) & \\ \cdashline{2-6}
 &  \multirow{2}{*}{7 TeV, $\mathscr{L}=$ 37 pb$^{-1}$}& inclusive jet  & \multirow{2}{*}{0.4 and 0.6} & ($ p_{\mathrm{T}}/\mathrm{GeV} \geq 20$) $\otimes~(|y| < 4.4$)   & \multirow{2}{*}{\citen{Aad:2011fc} }\\ 
 &  & dijet mass &  & for dijet mass: ($ p^{1st}_{\mathrm{T}}/\mathrm{GeV} \geq 30$) &  \\ \cdashline{2-6}
 & 7 TeV, $\mathscr{L}=$ 4.5 fb$^{-1}$ & inclusive jet &0.4 and 0.6&($ p_{\mathrm{T}}/\mathrm{GeV} \geq 100$) $\otimes~(|y| < 3.0$)  & \citen{Aad:2014vwa} \\ \cdashline{2-6}
 & 7 TeV, $\mathscr{L}=$ 4.5 fb$^{-1}$ & dijet mass &0.4 and 0.6&  ($p^{1st}_{\mathrm{T}}/\mathrm{GeV}  \geq 100$ and  $p^{2nd}_{\mathrm{T}}/\mathrm{GeV}  \geq 50$) $\otimes~(|y| < 3.0)$ & \citen{Aad:2013tea} \\ \hline
 &  \multirow{2}{*}{ 7 TeV, $\mathscr{L}=$ 2.9 pb$^{-1}$} &  \multirow{2}{*}{$\Delta \phi_{jj}$} &   \multirow{2}{*}{0.5} & ($ p_{\mathrm{T}}/\mathrm{GeV} \geq 30$) $\otimes~(|y| < 1.1$) &   \multirow{2}{*}{\citen{Khachatryan:2011zj}}\\
 & &  & & ($ p^{max}_{\mathrm{T}}/\mathrm{GeV} \geq 80$) & \\ \cdashline{2-6}
 \multirow{7}{*}{CMS}  &  7 TeV, $\mathscr{L}=$ 34 pb$^{-1}$& inclusive jet & 0.5& ($p_{\mathrm{T}}/\mathrm{GeV}  \geq 18$) $\otimes~(|y| < 3.0$) & \citen{CMS:2011ab}  \\ \cdashline{2-6}
     & 7 TeV, $\mathscr{L}=$ 36 pb$^{-1}$& dijet mass &0.7&($p^{1st}_{\mathrm{T}}/\mathrm{GeV}  \geq 60$ and  $p^{2nd}_{\mathrm{T}}/\mathrm{GeV}  \geq 30$) $\otimes~(|y| < 2.5)$   & \citen{Chatrchyan:2011qta} \\  \cdashline{2-6}
  &   \multirow{2}{*}{7 TeV, $\mathscr{L}=$ 5 fb$^{-1}$}& inclusive jet &   \multirow{2}{*}{0.7}& ($p_{\mathrm{T}}/\mathrm{GeV}  \geq 100$) $\otimes~(|y| < 2.5$)  & \multirow{2}{*}{\citen{Chatrchyan:2012bja}}\\ 
  &  & dijet mass &  &($p^{1st}_{\mathrm{T}}/\mathrm{GeV}  \geq 60$ and  $p^{2nd}_{\mathrm{T}}/\mathrm{GeV}  \geq 30$) $\otimes~(|y| < 2.5)$  &  \\ \cdashline{2-6}
    &\multirow{2}{*}{7 TeV, $\mathscr{L}=$ 5 fb$^{-1}$}  &  inclusive jet & \multirow{2}{*}{0.7} &(35 $\leq p_{\mathrm{T}}/\mathrm{GeV} < $150) $\otimes$ (3.2$<|\eta^{f}|<$4.7)  & \multirow{2}{*}{\citen{Chatrchyan:2012gwa}} \\ 
    & &  jet $p_{\mathrm{T}}$ & & in dijet evt.: ($|\eta^{c}|<$2.8)  &  \\   \cdashline{2-6}
    &7 TeV, $\mathscr{L}=$ 5 fb$^{-1}$  &  inclusive jet &0.5 and 0.7 & ($p_{\mathrm{T}}/\mathrm{GeV}  \geq 56$) $\otimes~(|y| < 3.0$)  & \citen{Chatrchyan:2014gia} \\  \cdashline{2-6}
  prelim. &  8 TeV, $\mathscr{L}=$ 10.71 fb$^{-1}$ &  inclusive jet &0.7 &  ($p_{\mathrm{T}}/\mathrm{GeV}  \geq 80$) $\otimes~(|y| < 3.0$)  & \citen{CMS:2013kda,CMS:2013tea} \\  \cdashline{2-6}
  prelim. &  8 TeV, $\mathscr{L}=$ 9.2 fb$^{-1}$ &  dijet mass &0.7 &  & \citen{CMS:2014aga} \\  \cdashline{2-6}
  prelim. &  8 TeV, $\mathscr{L}=$ 19.7 fb$^{-1}$ & $\Delta \phi_{jj}$ &0.7 &  & \citen{CMS:2015aya} \\  \cdashline{2-6}
  prelim. &  2.76 TeV, $\mathscr{L}=$ 5.43 pb$^{-1}$ & inclusive jet  &0.7 &  & \citen{CMS-PAS-SMP-14-017}\\ \hline \hline
\end{tabular}
}
\label{table:paper}
}
\end{table}

\clearpage
\newpage
\section{Jet reconstruction and calibration}
\label{sec:jetreco}
A crucial part of the jet cross section measurements is the measurement of the direction and the momentum of the jet .
Given the different detector technologies, ALICE, ATLAS and CMS used different techniques to measure jets\cite{Schwartzman:2015ada}. 

In ATLAS, the input objects to the jet algorithm are three-dimensional “topological” clusters from calorimeter cells.

Taken as input to the anti-$k_{\mathrm{t}}$ jet  algorithm, each cluster is considered as a massless particle with an energy  equal to the sum of the energy in its cells, and a direction given by the energy-weighted barycenter of the cells in the cluster with respect to the geometrical centre of the ATLAS detector.
The four-momentum of a non-calibrated jet is defined as the sum of four-momenta of the clusters making up the jet.

In CMS, the inputs to the jet clustering algorithm are the four-momentum vectors of the reconstructed particle candidates. Each such candidate is constructed using the particle-flow technique, which combines the information from several sub-detectors.

For both ATLAS and CMS, the reconstructed jets require additional energy corrections to mostly account for the non-linear and non-uniform response of the calorimetric system to hadrons. They can be summarized as:
\begin{itemize}
\item Subtraction of the extra energy due to additional proton-proton collisions within the same or neighboring bunch crossings; 
\item Correction of the energy of the jets using Monte Carlo (MC) simulations;
\item An additional in-situ calibration is applied to correct for residual differences between MC simulation and data, derived by combining the results of $\gamma$--jet, $Z$--jet, dijet and multijet momentum balance techniques.
\end{itemize}

Once these calibrations are applied, the reconstructed jets \pT~are corrected to the hadron level on average. 
The typical jet \pT~calibration systematic uncertainty (usually defined as jet energy scale uncertainty) is of the order of 1\%  at \pT~= 100 GeV for both ATLAS and CMS.
The typical jet \pT~resolution is 10--15\% at \pT~= 100 GeV. 
To correct the measured distributions for the effect of the finite resolution, dedicated strategies, called unfolding, have been used by the different Collaborations, and are discussed in Section \ref{sec:systematics}.

\section{Trigger and event selection}
\label{sec:eventsel}

 The large cross section predicted for the jet production caused a very strong interest in performing the measurements already with the very first LHC collisions.  During the first LHC collisions, most of the experimental techniques, including the trigger, were under commissioning, so the measurements of these cross sections were interesting per-se, and as a natural test-bench for most of the detector performances.

At the very beginning of the data taking, very simple trigger systems have been used to collect the data.
This is the case, for example, of the minimum-bias trigger scintillators (MBTS) in ATLAS, consisting of 32 scintillator counters 

covering the range of 2.09 $<|\eta|<$3.84\footnote{The minimum-bias triggers have been used to measure the jet cross sections in the low \pT~region, where the the jet triggers are not fully efficient.}.
 Thanks to the increased confidence gained by studying the trigger performance in the first collisions, calorimeter based triggers have been  fully commissioned and used  to collect data for the jet cross section measurements in the rest of the Run 1. 

For both ATLAS and CMS the trigger chain is split in different steps:
\begin{itemize}
\item a first fast custom-built hardware decision (called Level 1, or L1);
\item a second more refined software decision (High Level Trigger - HLT for CMS, which is split in Level 2 and Event Filter for ATLAS).
\end{itemize}
The L1-jet reconstruction uses  electromagnetic and hadronic calorimeter towers with a specific granularity of $\Delta\phi\times\Delta\eta$.  L1-jets are formed by grouping these towers, which contribute to define the transverse energy (\ET) of the L1-jet.  
The single jet trigger selects the event if one or more L1-jet passes a predetermined threshold in \ET. 
In the software trigger step(s), this decision is refined by measuring in a more accurate way the properties of the jets identified at L1. The event is selected if at each step of the single jet trigger chain, at least one of the  jets passes a predetermined \ET~threshold .

With the increase of the LHC instantaneous luminosity, the rate of events having a jet passing a certain chain of \ET~thresholds in the jet trigger increased. To maintain the overall rate of the data acquisition in its limits, only single jet trigger chains with sufficiently high \ET~thresholds could be used. To extend the measurement to lower jet \pT, trigger chains with lower \ET~ thresholds have been randomly enabled with an event by event stochastic decision, named pre-scale. 
As an example, with a pre-scale of $p$, a trigger chain is enabled only for $1/p$ of the total events, reducing the rate by a factor $p$.

Given the evolution of the trigger conditions and pre-scales, the LHC Collaborations designed a strategy to get the smallest statistical uncertainty in each region of the measurements, while keeping the potential trigger bias as small as possible.

\begin{figure}
\includegraphics[width=0.49\textwidth]{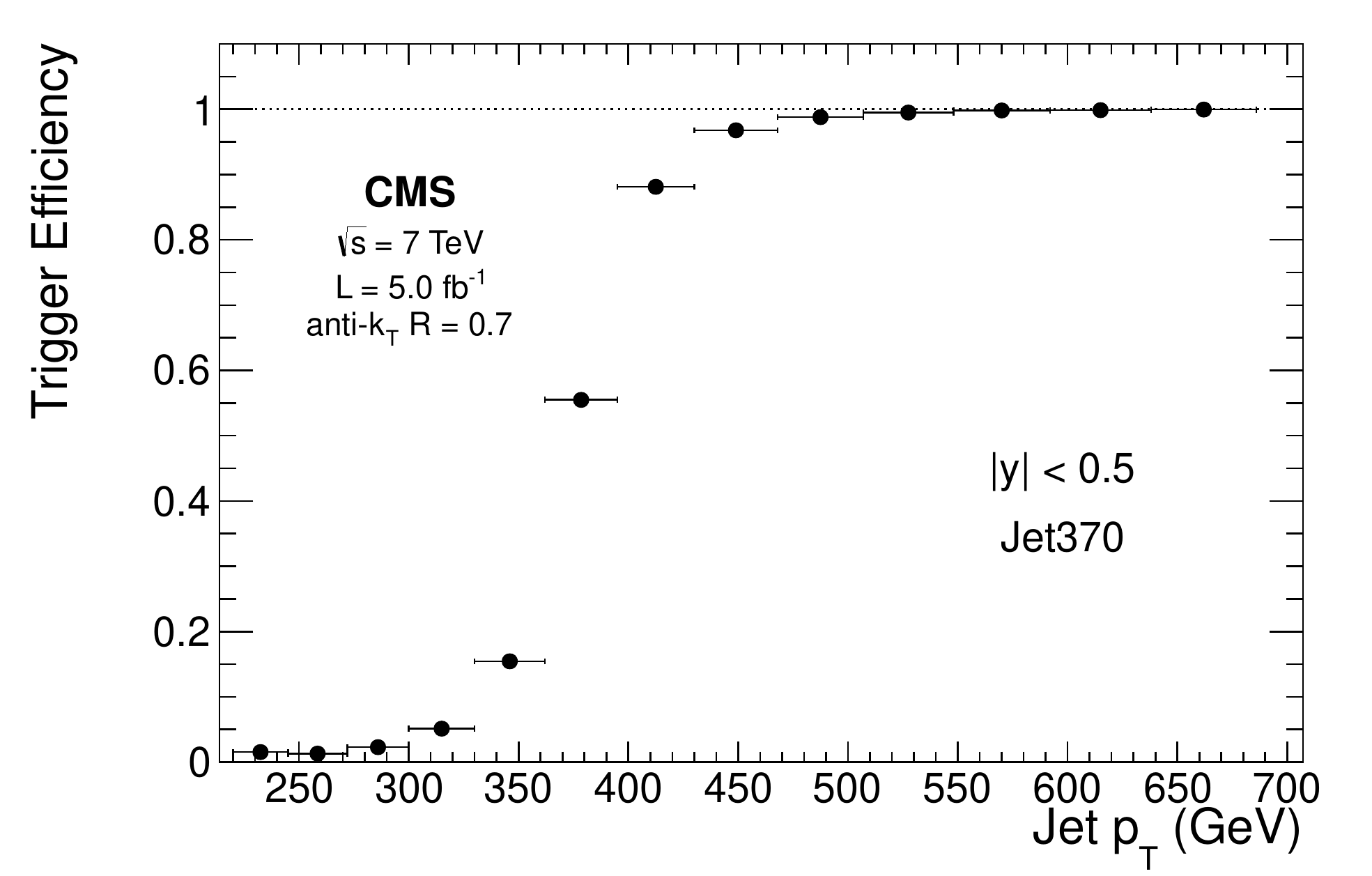}
\includegraphics[width=0.49\textwidth]{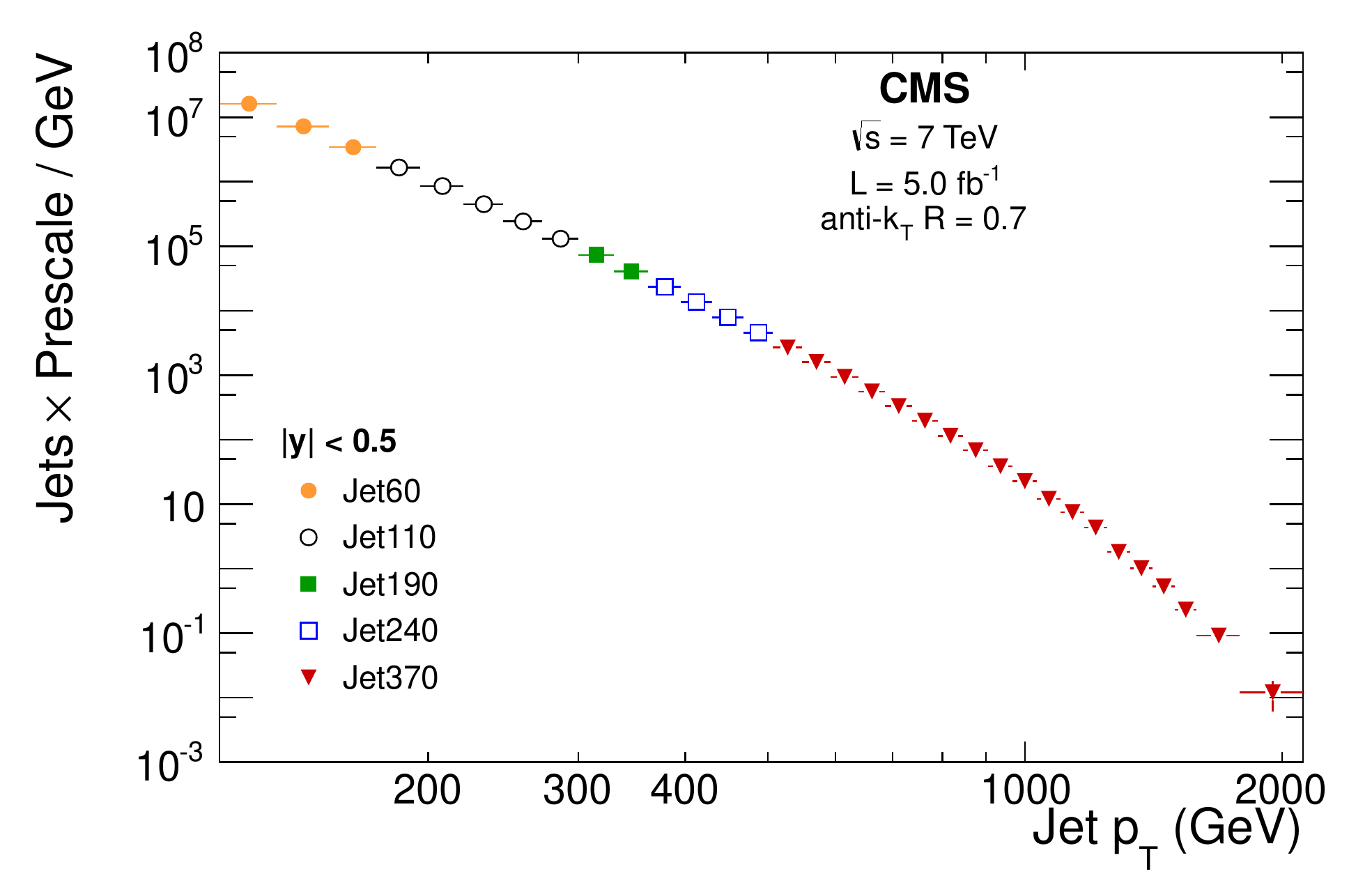}
\caption{\label{fig:singlejetTrigger}
Left: Trigger efficiency as a function of the jet \pT~for the 370 GeV single-jet trigger and for the central rapidity bins.
Right: Spectrum construction from individual trigger paths for the inclusive jet \pT~spectrum for $|y| < 0.5$. Taken from Ref. \citen{Chatrchyan:2012bja}
}
\end{figure}

Fig. \ref{fig:singlejetTrigger} shows an examples of the single jet trigger efficiency  as measured by CMS 
 using events collected with a lower threshold single-jet trigger (and confirmed with events collected with single-muon
triggers), and the final distribution as measured by each trigger chain.

The selected events are additionally cleaned from periods with detector problems, and 
from events with non-physical jets, originated by cosmic rays, beam background, or noise in the detector.

All the jets or events selected by the proper trigger chain and passing the quality criteria, have been used for the measurement of the inclusive jet cross section, the dijet cross section and the dijet azimuthal decorrelation. The next Section will discuss the strategy used to correct for the detector effects which are still present in the spectrum shown by the plot on the right in Fig.  \ref{fig:singlejetTrigger}.

\section{Unfolding and experimental systematic uncertainties}
\label{sec:systematics}

The most important detector effects present in the spectrum on the right in Fig.  \ref{fig:singlejetTrigger} are:
\begin{itemize}
\item The efficiency in measuring a jet $\epsilon$;
\item The purity in measuring a real jet (and not a fake from detector instabilities) $\pi$;
\item The effect of the finite resolution in the measurement of the energy and direction of the jet.
\end{itemize}
The first two effects are usually very small: the ATLAS and CMS detectors  have very high efficiency in measuring a jet with \pT~$>$~20 GeV, and several cleaning cuts have been adopted to avoid spurious jets due to detector or beam background conditions.

The effect of the finite resolution plays a more important role in these measurements.
In the inclusive jet cross section measurement the effect of the smearing due to the energy resolution 
causes migrations between the bins of the measurement.  Due to the steeply falling spectrum, more events migrate into the bin, than out.
In order to allow for a direct comparison of experimental measurements with corresponding results from other experiments and with theoretical predictions, special corrections, named unfolding corrections, are applied. 

In general, the number $n_{i}$ of jets (or events) in a bin $i$ of the unfolded jet cross sections can be expressed as:
\begin{equation}
n_{i}=\Sigma_{j} \frac{1}{\epsilon_{i}}\times A_{i,j}\times\pi_{j} \times m_{j}=\Sigma_{j} B_{i,j} \times m_{j}
\end{equation}
where $m_{j}$ is the number of jets (or events) measured in the bin $j$, $\pi_{j}$ is the purity of the jets (or events) measured in the bin $j$, $A_{i,j}$ is the response matrix that associates jets (or events) measured in the bin $j$ with the estimated number of jets (or events) in the unfolded bin $i$, and finally,  $\epsilon_{i}$ is the efficiency in measuring a jet (or selecting an event) for the bin $i$. 

Different techniques to build $B_{i,j}$ have been used in Run1. The very first measurements at the LHC used a bin-by-bin correction, in which $B_{i,j}=k_{j}\delta_{i,j}$, and $k_{j}$ is determined by the ratio of the reconstructed spectrum and the hadron level spectrum in Monte Carlo simulations. 
More advanced techniques such as the iterative bayesian method\cite{D'Agostini1995487} or the Iterative, Dynamically Stabilised (IDS)\cite{2009arXiv0907.3791M} method  have been tested and used already at the end of the first year of the data taking in Run1.
The benefits of these advanced techniques are the use of the full information on the response matrix which allows the propagation of the statistical and systematic uncertainties on the final measurement.

\begin{figure}
\includegraphics[width=0.55\textwidth]{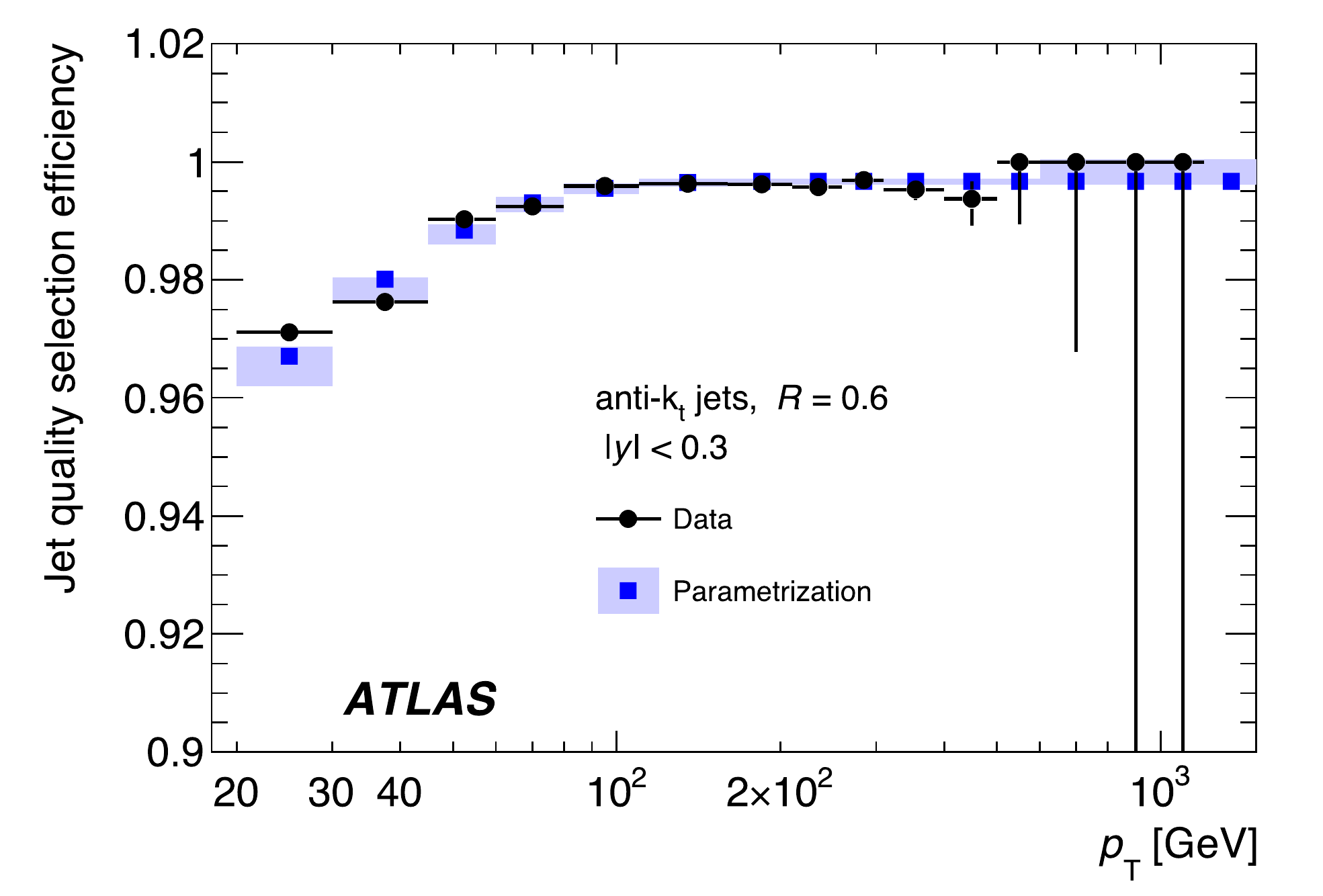}
\includegraphics[width=0.39\textwidth]{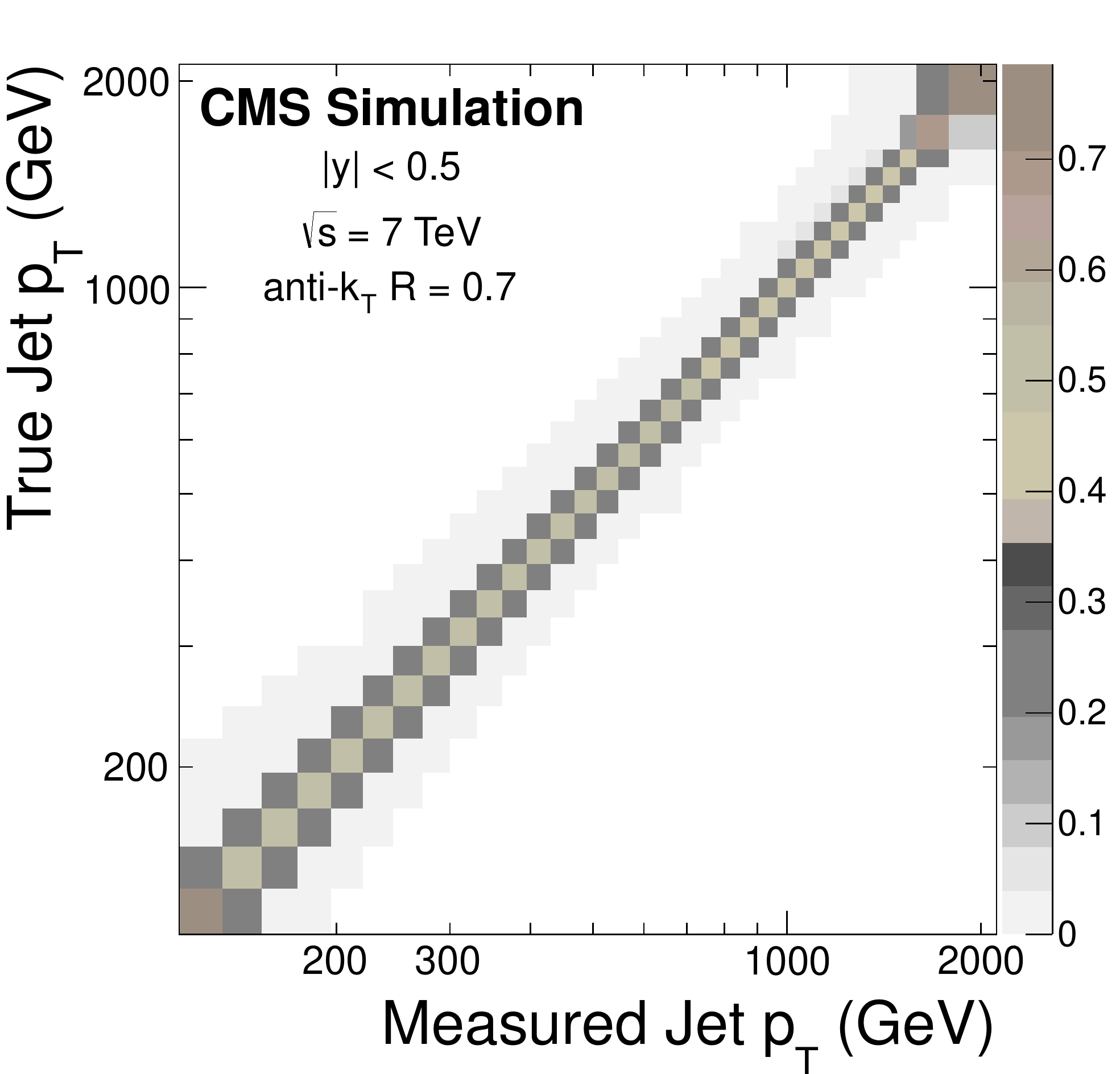}
\caption{\label{fig:Unfolding}
Left: Efficiency $\epsilon$ for jet quality selection as a function of \pT.
The black circles indicate the efficiency measured in-situ using a tag-probe method. 
Taken from Ref. \citen{Aad:2011fc}.
Right: Response matrix for the inclusive jet \pT~spectrum.
Taken from Ref. \citen{Chatrchyan:2012bja}
}
\end{figure}

As an example, Fig. \ref{fig:Unfolding} shows the $\epsilon_{i}$ as estimated by ATLAS for the inclusive jet cross section, and the response matrix estimated by the CMS collaboration. 

\begin{figure}
\includegraphics[width=0.495\textwidth]{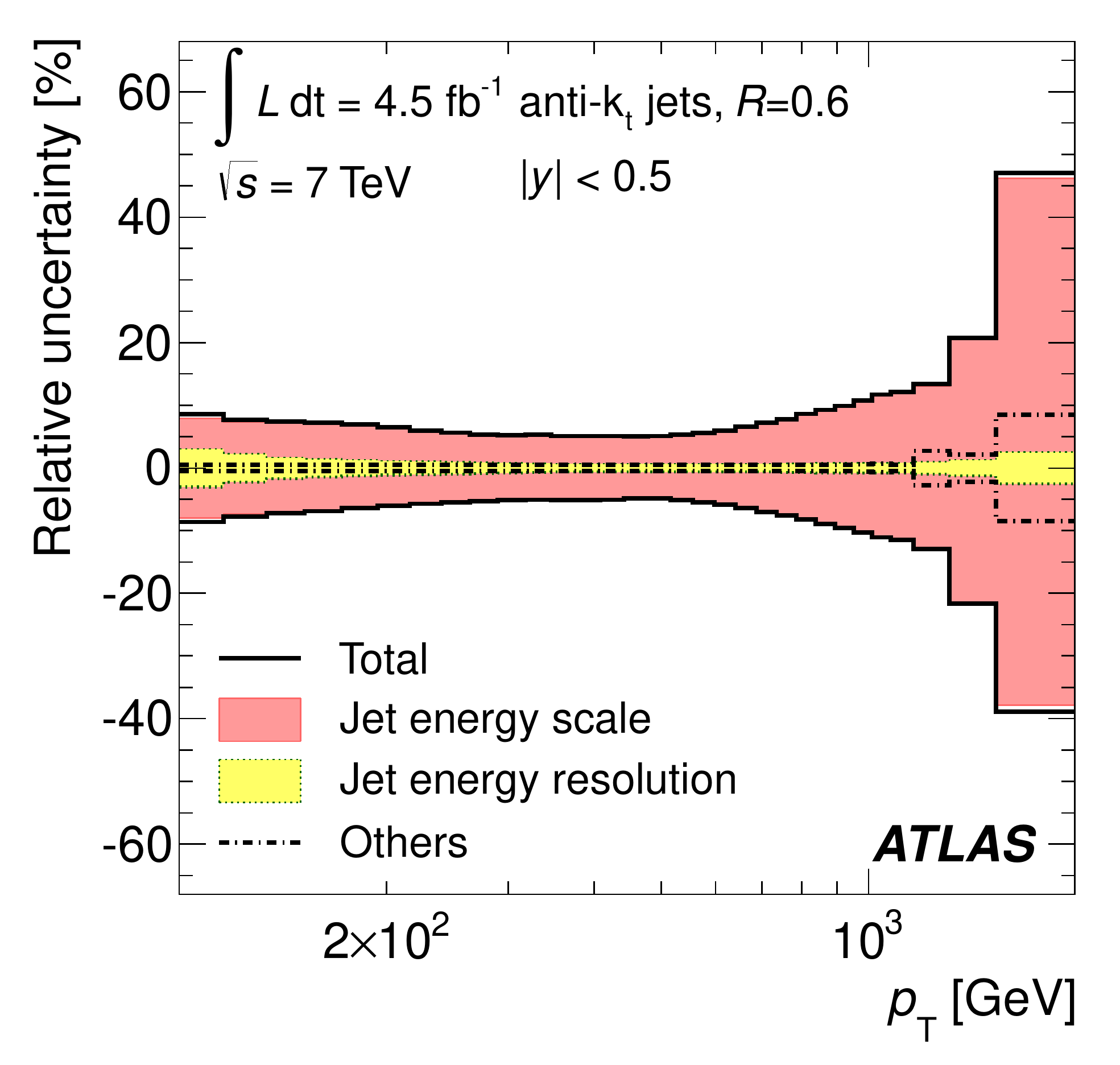}
\includegraphics[width=0.495\textwidth]{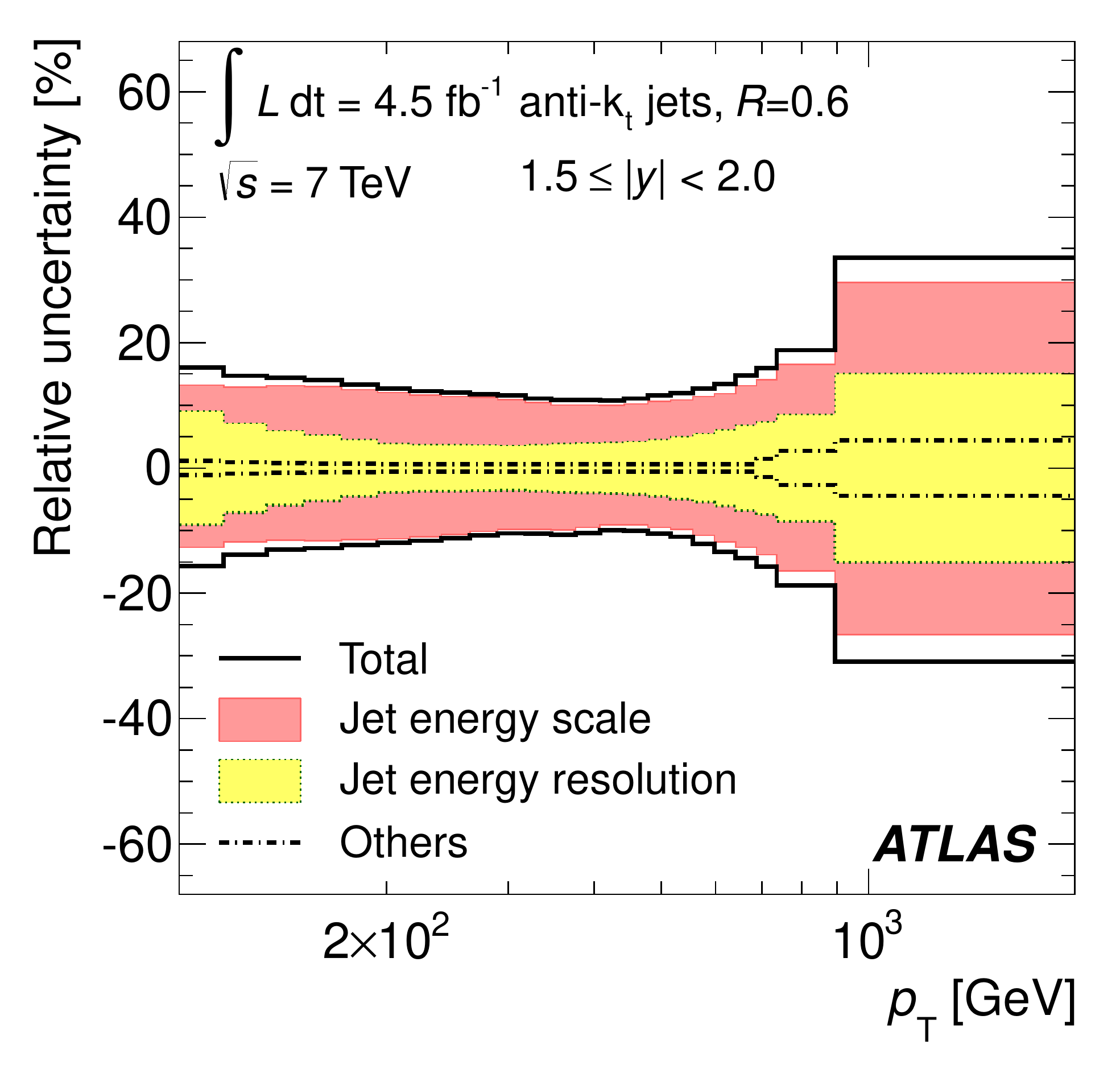}
\includegraphics[width=0.495\textwidth]{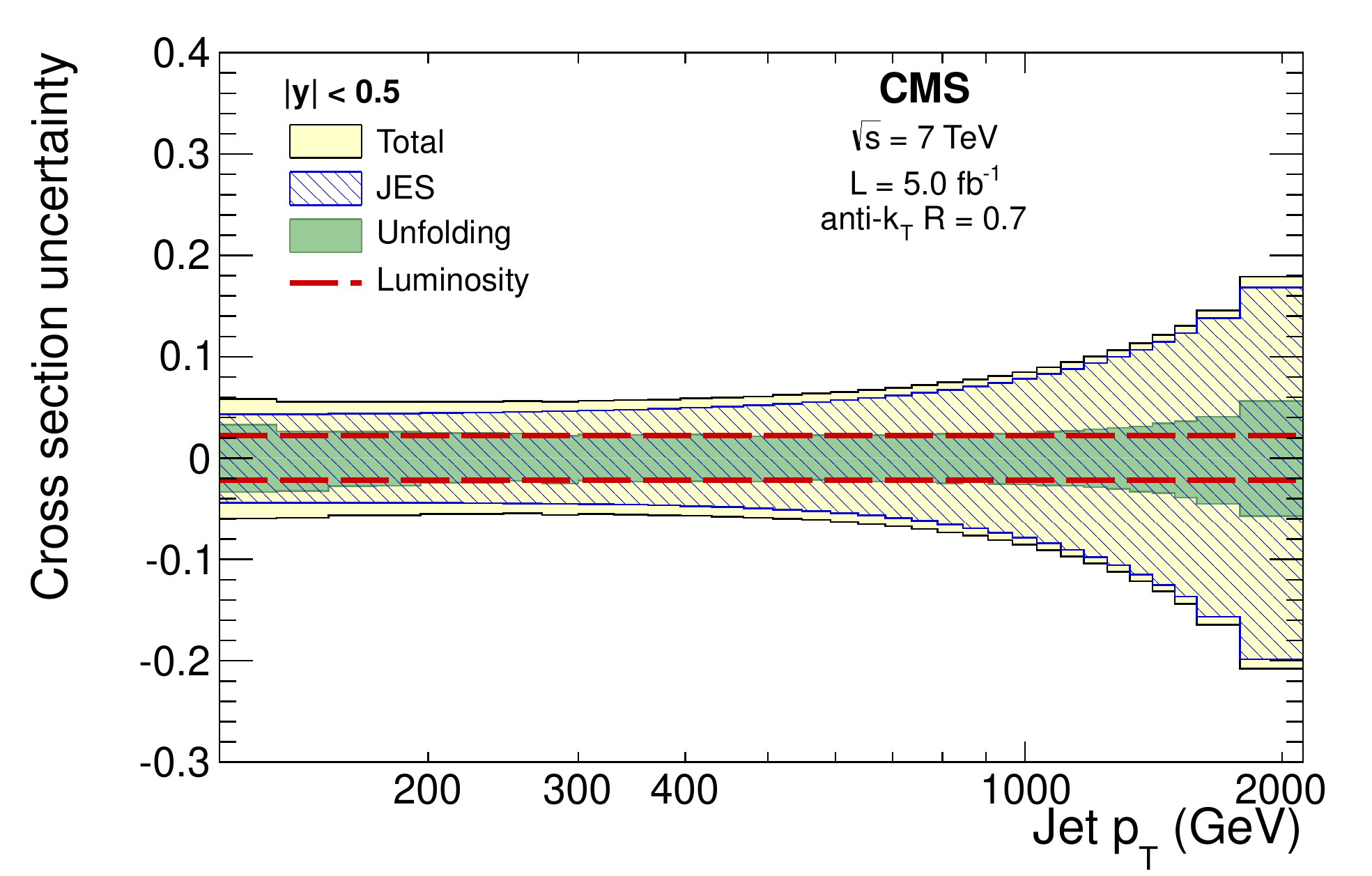}
\includegraphics[width=0.495\textwidth]{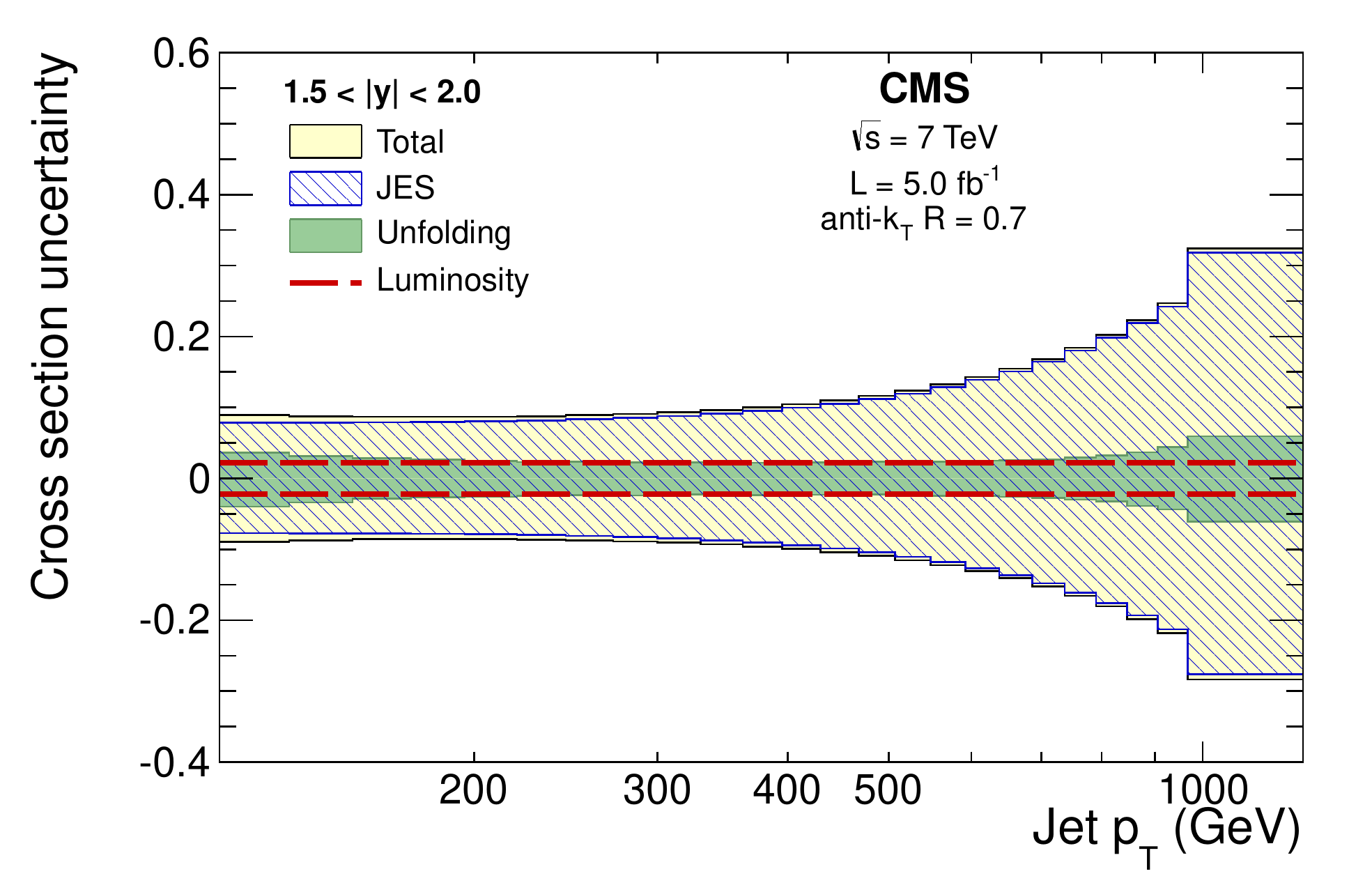}
\caption{\label{fig:Syst}
Experimental systematic uncertainties in the inclusive jet cross section measurement in two representative rapidity bins, as a function of the jet \pT. In addition to the total uncertainty, the uncertainties from the jet energy scale (JES), the jet energy resolution (JER) and other systematic sources are shown separately for both ATLAS (first row, taken from Ref. \citen{Aad:2014vwa}) and CMS (second row, taken from Ref. \citen{Chatrchyan:2012bja}).
}
\end{figure}

The final ingredient for the measurement of the jet cross section is the determination of the systematic uncertainties. The dominant experimental uncertainties are due to the lack of knowledge 
of the jet energy scale (JES), the luminosity, and the jet \pT~resolution (JER). Other sources of systematic uncertainty, such as the jet angular resolution, the unfolding uncertainties and the trigger inefficiencies are usually much smaller. 
Fig. \ref{fig:Syst}  shows their impact on the inclusive jet cross section in two representative regions in rapidity for both ATLAS and CMS. The plots give a clear indication that the dominant systematic uncertainty is  coming from the jet energy scale uncertainty.
Due to the shape of the spectrum, an uncertainty of 1--2\% on the calibration of the jets  leads to an uncertainty of  5--10\% in the inclusive jet cross section.

\section{Theoretical predictions}
\label{ThePred}
The theoretical predictions for the inclusive and dijet cross sections have seen important improvements in the last years.
While before the LHC era, the measured jet cross sections were usually compared with theoretical predictions obtained from next to leading order (NLO) pQCD calculations (for example using the \nlojetpp\cite{Nagy:2003tz}  program) with corrections for non-perturbative effects,
in the last years, predictions from NLO matrix elements interfaced to a Monte Carlo simulation of parton showers (PS), hadronisation and underlying event, which are relevant for the measurements of the jet cross sections with different parameters $R$, bacame available (for example using \powheg\cite{Alioli:2010xd}). 

In addition, corrections for electroweak effects\cite{Dittmaier:2012kx} have been used in the comparison of the measured jet cross sections (an example is discussed in Ref. \citen{Aad:2014vwa}).
These corrections comprise tree-level effects of $O(\alpha\alpha_\mathrm{S},\alpha^2)$ as well as weak loop effects of $O(\alpha\alpha_\mathrm{S}^2)$ on the cross section, where $\alpha$ is the electroweak coupling constant.

The correction reaches more than 10\% for \pT~$>1$~TeV in the lowest rapidity
bin, but decreases rapidly as the rapidity increases. It is less than 1\% for jets with $|y|>1$.

These improvements open the possibility to compare the measured cross sections both to NLO pQCD calculations corrected for non-perturbative and electroweak effects and to NLO+PS MC generated events and electroweak corrections.
Thanks to fast and flexible interfaces to the  \nlojetpp~calculation developed in the last years, such as APPLGRID\cite{applgrid:2009} and the fastNLO\cite{Kluge:2006xs},  NLO pQCD calculations with various parton density function (PDF) sets, various values of $\alpha_s$ and various values of the renormalization $\mu_\mathrm{R}$ and factorization $\mu_\mathrm{F}$ scales have become very  CPU efficient.
This has the clear advantage of helping in detailed studies of the stability of the theoretical prediction under variations of PDF, \alphas, $\mu_\mathrm{R}$ and  $\mu_\mathrm{F}$ and in the use of the measured cross sections  to extract the information of the underlying dynamics.

All these improvements in the theoretical predictions are leading  the jet cross section measurements to become more and more important for a precise understanding 
of the dynamics of the interactions in high energy collisions.

To give an example for  \pT~=~1 TeV and $|y|<0.5$, the theoretical predictions on the inclusive jet cross section have an uncertainty of 4--10\% when  varying the renormalisation and factorisation scales; an uncertainty of the order of 5--10\% due to PDF errors, 2\% due to the error on \alphas, 2--3\% for the non perturbative effects.

\section{Measurements}	
\label{sec:measurements}

\subsection{Inclusive jet cross section}
\begin{figure}[h!]
\centering
\includegraphics[width=0.495\textwidth]{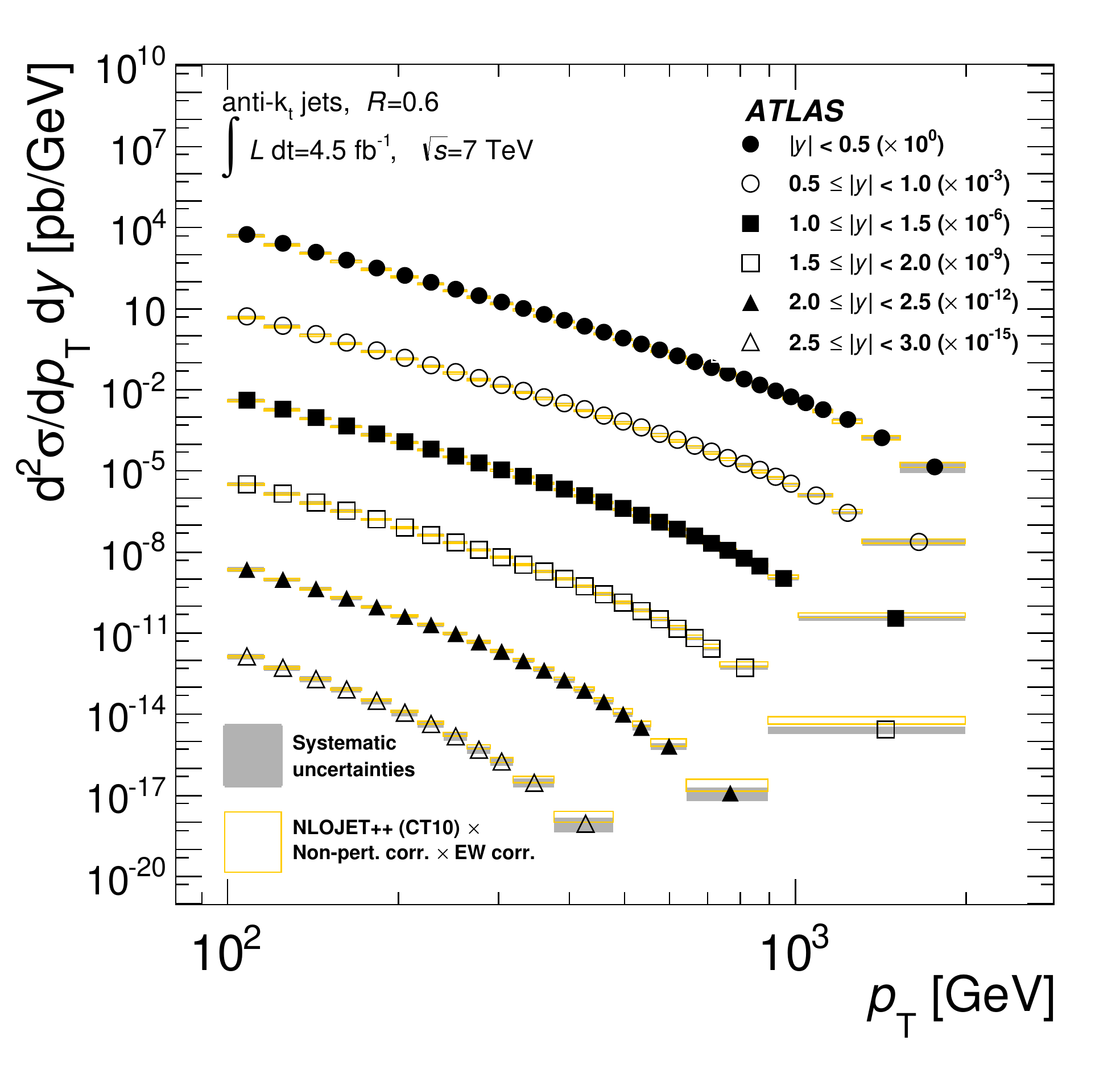}
\includegraphics[width=0.495\textwidth]{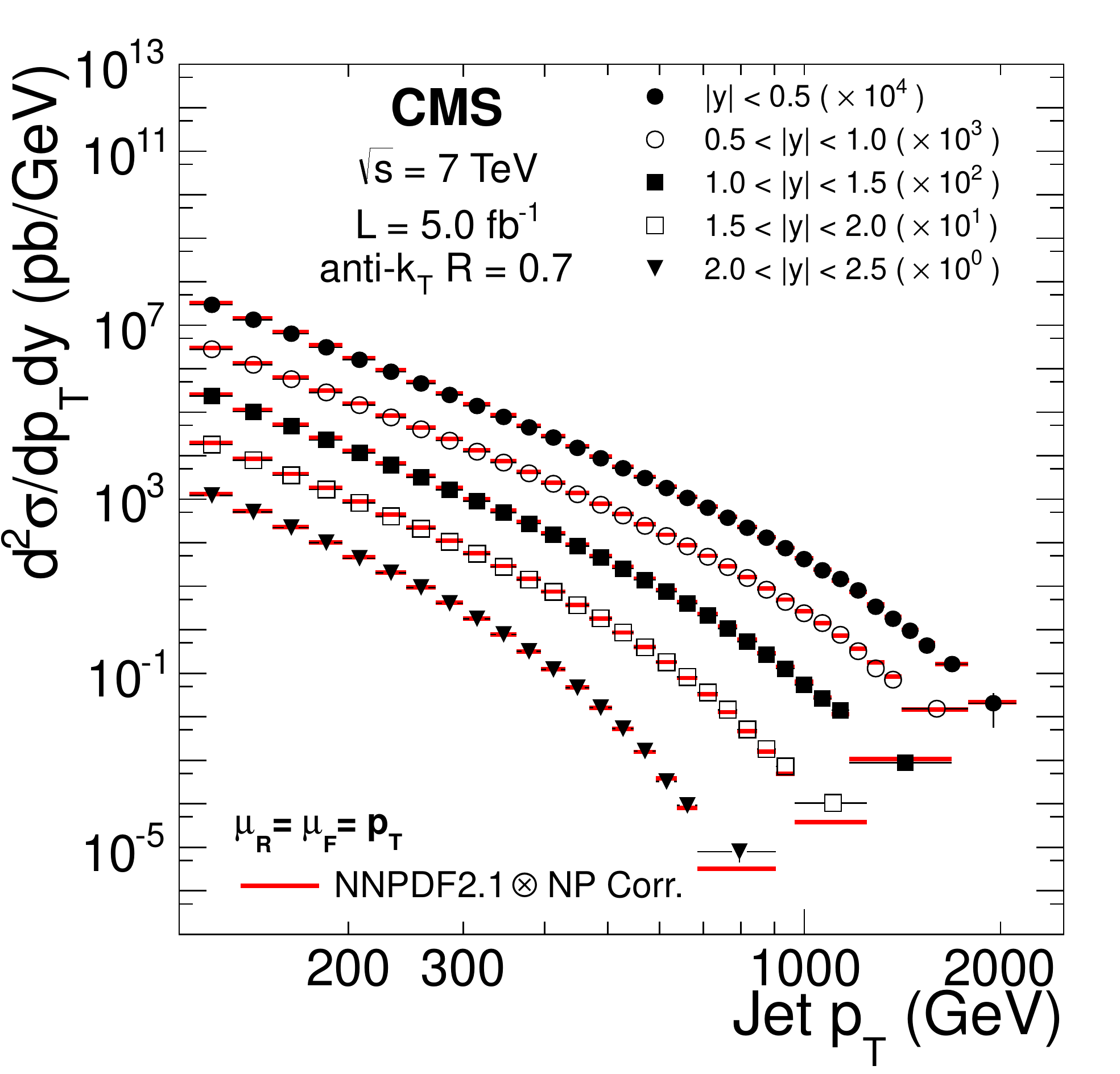}
\caption{\label{fig:ATLAS7TeV}
Double-differential inclusive jet cross sections as a function of the jet \pT~in bins of rapidity, for anti-$k_{\mathrm{t}}$ jets with $R$=0.6 measured by the ATLAS collaboration on the left and with $R$=0.5  measured by the CMS collaboration on the right. For presentation, the cross sections are multiplied by the factors indicated in the legend. The data are compared to NLO pQCD predictions calculated using \nlojetpp~with the CT10 NLO PDF for the measurement performed by ATLAS  and NNPDF2.1 for the measurement by CMS, to which non-perturbative corrections and electroweak corrections (for ATLAS) are applied.
Taken from Ref. \citen{Aad:2014vwa} and \citen{Chatrchyan:2012bja}.
}
\end{figure}

The double-differential inclusive jet cross sections for $\sqrt{s}=7$ TeV measured by the ATLAS and the CMS collaborations are shown in Fig.~\ref{fig:ATLAS7TeV}. 
These measurements extend over jet transverse momenta from 100~GeV to 2~TeV in the rapidity region $|y|<3.0$ (ATLAS) and $|y|<2.5$ (CMS)\footnote{Previous measurements extend to lower \pT~and higher $|y|$. See Table \ref{table:paper} for a complete list.}. 
The NLO pQCD predictions calculated with \nlojetpp~with corrections for non-perturbative effects and electroweak effects (for ATLAS) applied are compared to the measurement. 
The figure shows that the NLO pQCD predictions reproduce the measured cross sections, which range over eight orders of magnitude in the rapidity bins.

\begin{figure}
\includegraphics[width=0.495\textwidth]{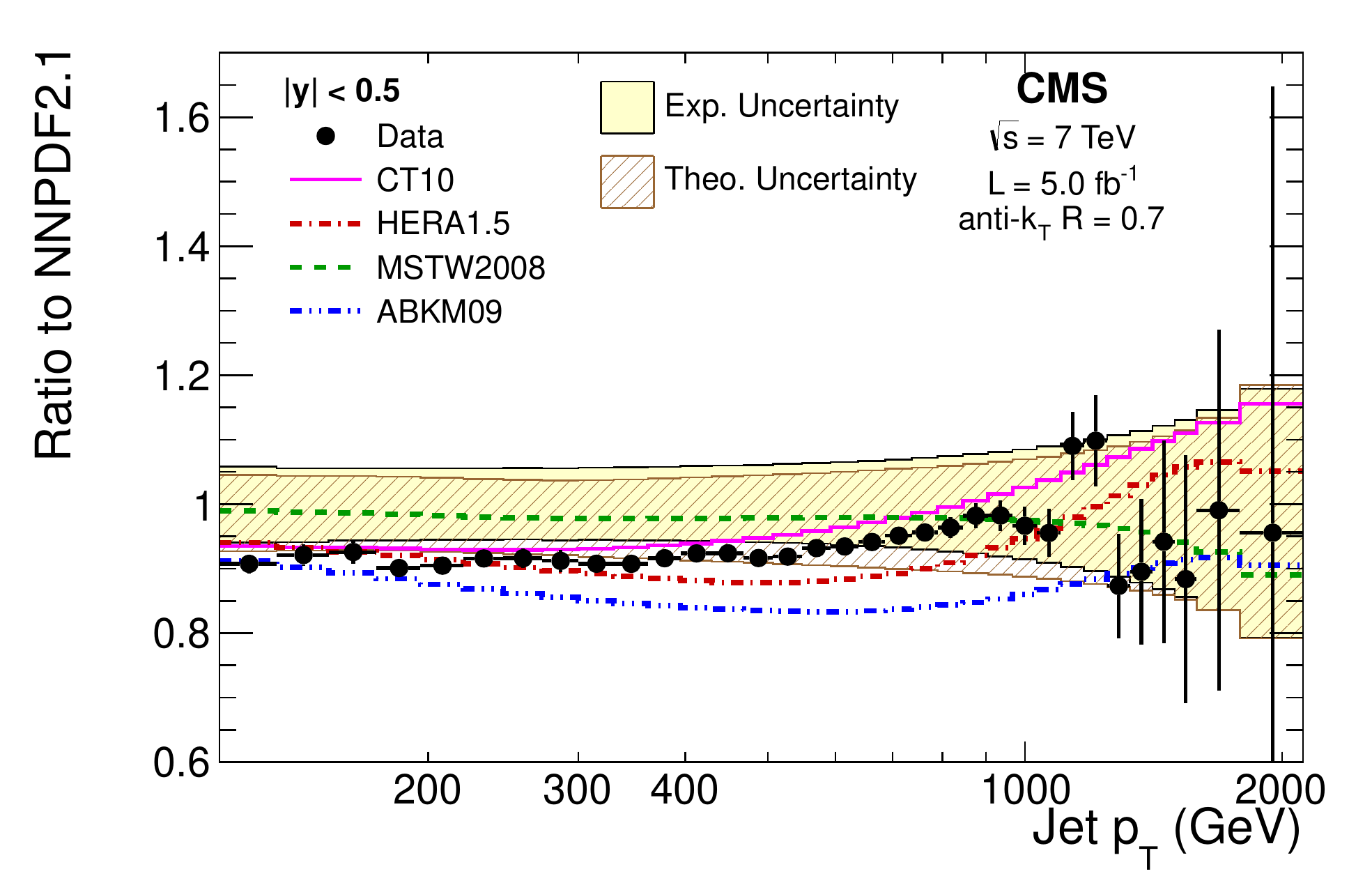}
\includegraphics[width=0.495\textwidth]{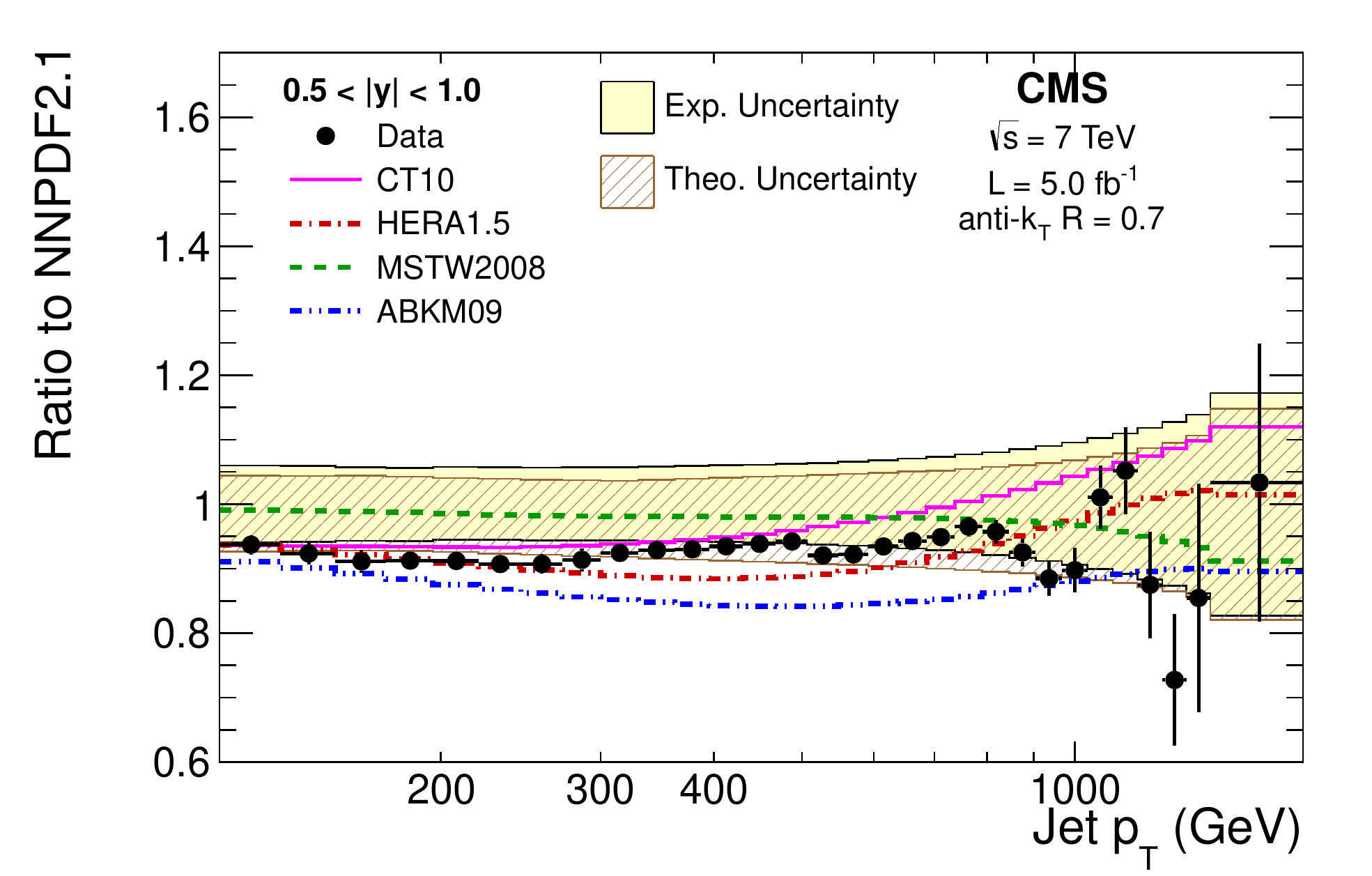}
\caption{\label{fig:PDFCMS}
Ratio of inclusive jet cross sections to the theoretical prediction using the central value of the NNPDF2.1 PDF set for a central $|y|$ bin. The solid histograms show the ratio of the cross sections calculated with the other PDF sets to that calculated with NNPDF2.1. The experimental and theoretical systematic uncertainties are represented by the continuous and hatched bands, respectively.
Taken from Ref. \citen{Chatrchyan:2012bja}.
}
\end{figure}

Both collaborations performed more detailed comparisons with different PDF sets, and with NLO+PS MC. As an example, Fig. \ref{fig:PDFCMS} shows the ratio
 of the inclusive jet cross sections to the theoretical prediction using the central value of the NNPDF2.1 PDF~\cite{Ball:2010de,Forte:2010ta} set. The plots show the effect of calculating the cross section with other PDF sets, like CT10~\cite{Lai:2010vv}, MSTW2008~\cite{Martin:2009iq}, ABM 09 ~\cite{Alekhin:2012ig} and HERAPDF 1.5~\cite{HERAPDF15}, showing that all the predictions are generally in agreement with the measured cross section.

\begin{figure}
\includegraphics[width=0.99\textwidth]{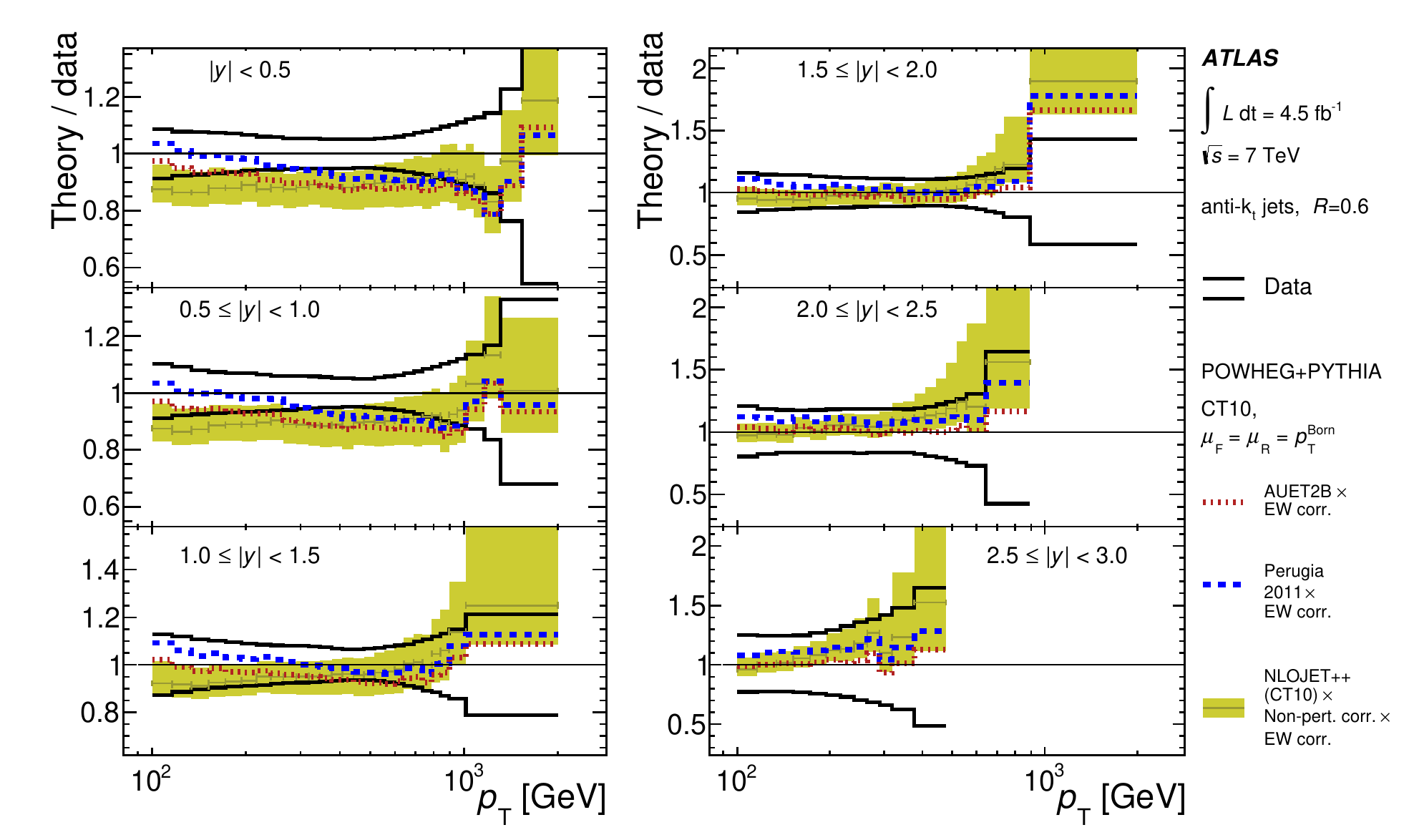}
\caption{\label{fig:TunesATLAS}
Ratio of predictions from \powheg~to the measured double-differential inclusive jet cross section, shown as a function of the jet \pT~in bins of jet rapidity. The figure also shows the NLO pQCD prediction using \nlojetpp~with the CT10 NLO PDF set, corrected for non-perturbative effects and electroweak effects. The \powheg~predictions use \pythia~for the simulation of parton showers, hadronisation, and the underlying event with the AUET2B tune and the Perugia 2011 tune. Electroweak corrections are applied to the predictions. Taken from Ref. \citen{Aad:2014vwa}.
}
\end{figure}

Fig. \ref{fig:TunesATLAS} shows how the MC NLO+PS predictions performed with \powheg+\pythia~compare with the measured data. In this comparison, performed by the ATLAS collaboration, two different tunes are considered: AUET2B~\cite{ATL-PHYS-PUB-2011-009} and Perugia 2011~\cite{Skands:2010akv4}. 
Both calculations are in agreement with the pQCD NLO calculation corrected for non-perturbative and electroweak  effects, but they show a better agreement with data, especially for \pT~$<$~500 GeV in the central rapidity bins, highlighting the necessity of a coherent treatment of perturbative and non perturbative dynamics in the calculation of the inclusive jet cross section.

\begin{figure}
\includegraphics[width=0.49\textwidth]{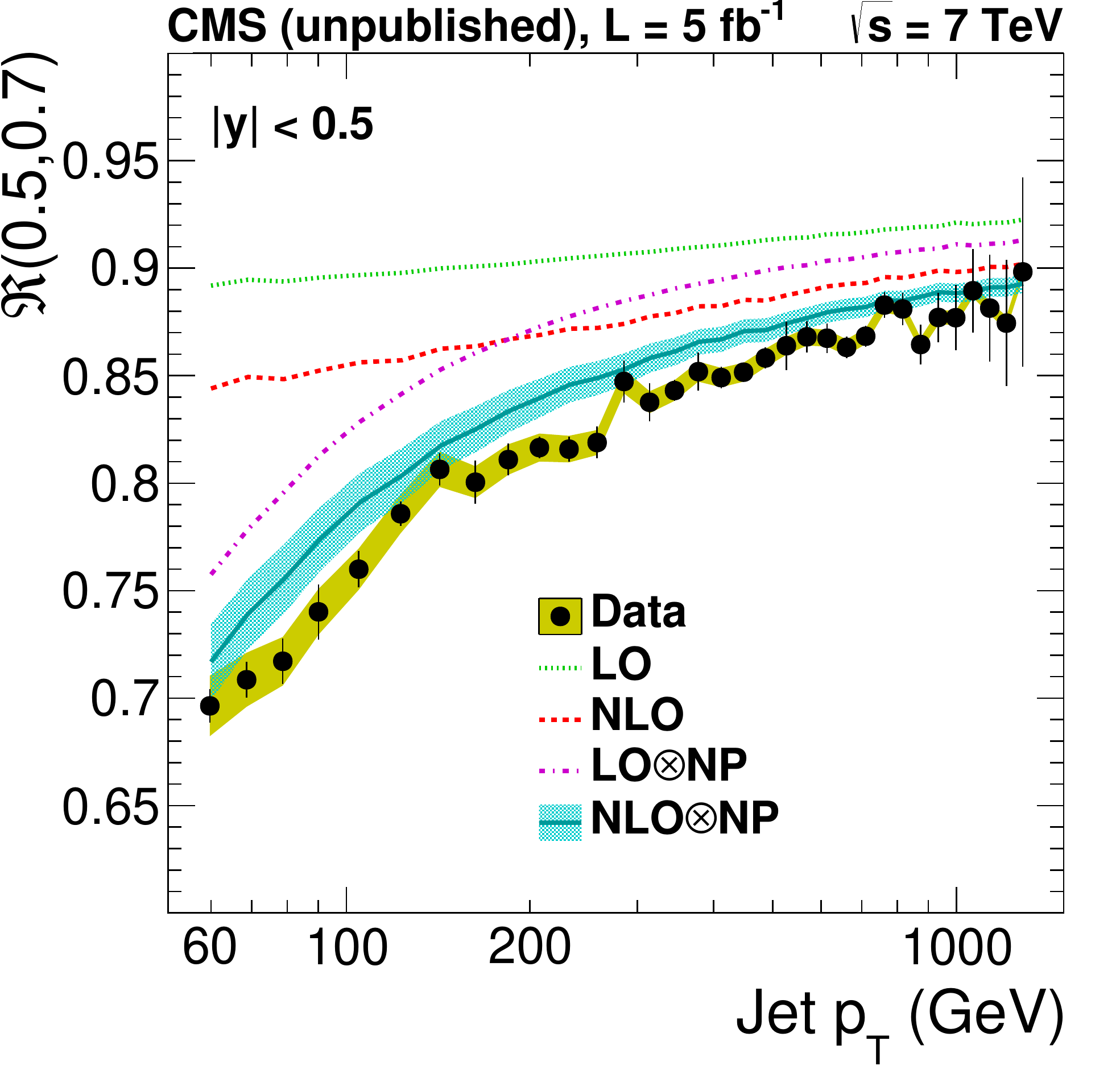}
\includegraphics[width=0.49\textwidth]{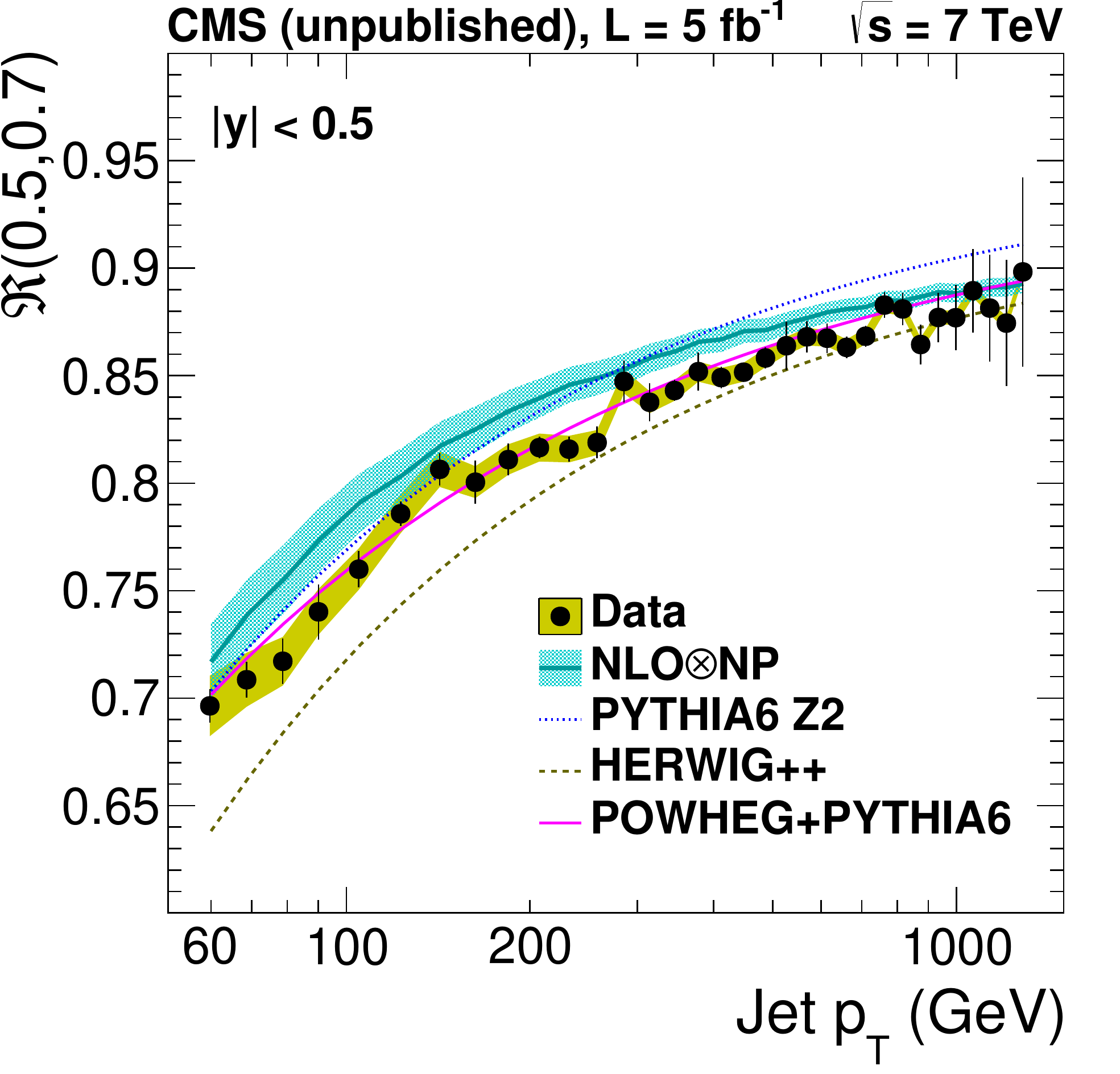}
\caption{\label{fig:CMSRatio}
Jet radius ratio R(0.5, 0.7) for $|y| < 0.5$, compared to LO and NLO with and without non-pertrurbative corrections (left) and versus NLO+PS and MC predictions (right). The error bars on the data points represent the statistical and uncorrelated systematic uncertainty added in quadrature, and the shaded bands represent correlated systematic uncertainties. 
Taken from Ref. \citen{Chatrchyan:2014gia}.
}
\end{figure}

An additional example of the interplay  of perturbative and non-perturbative dynamics is given in Ref. \citen{Chatrchyan:2014gia}. In this case, by measuring the ratio of the cross section with different $R$ parameters ($R$=0.5 and $R$=0.7), one gets a very powerful handle to investigate the perturbative and non-perturbative effects.
Although the cross sections themselves can be described within the theoretical and experimental uncertainties by predictions of pQCD at NLO, this is not the case for the ratio as shown in Fig. \ref{fig:CMSRatio}.  NLO calculations, even when complemented with nonperturbative corrections, are in clear disagreement with the data, while the MC event generators  \pythia~and \herwigpp~are in better accord with the measured jet radius ratio. The best description of this ratio is obtained by matching the cross section prediction at NLO with parton showers, as studied here using  \powheg+\pythia~for the showering, underlying event, and hadronization parts.

\begin{figure}
\includegraphics[width=0.99\textwidth]{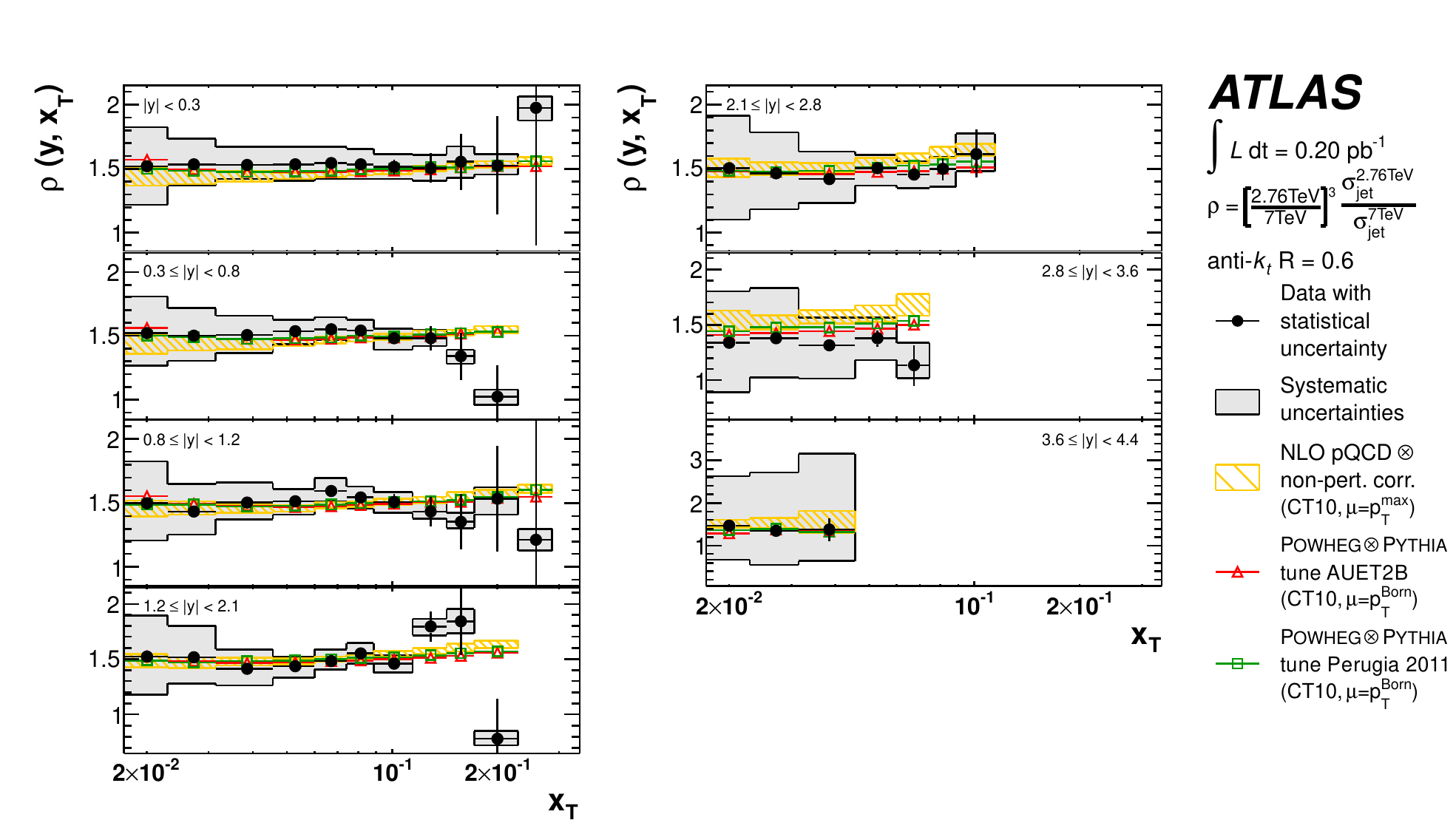}
\caption{\label{fig:ATLAS}
Ratio of the inclusive jet cross section at $\sqrt{s}=$  2.76 TeV to the one at $\sqrt{s}=$ 7 TeV as a function of $x_{\mathrm{T}}$ in bins of jet rapidity, for anti-$k_{\mathrm{t}}$ jets with $R$ = 0.6. 
Taken from Ref. \citen{Aad:2013lpa}.
}
\end{figure}

Another interesting ratio of cross sections, performed in Ref.  \citen{Aad:2013lpa}, is shown in Fig. \ref{fig:ATLAS}. In this case, the ratio is performed for measurements of the inclusive jet cross section at $\sqrt{s}=2.7$ TeV and $\sqrt{s}=7$ TeV. The measurement is found to be almost constant as a function of $x_{\mathrm{T}}~=~2$\pT$/\sqrt{s}$. This approximately constant behavior reflects both the asymptotic freedom of QCD and evolution of the gluon distribution in the proton as a function of the QCD scale. When measured as a function of \pT~, the cancellation of some of the systematic uncertainty in the ratio suggests that the measured jet cross section may contribute to constrain the PDF uncertainties. 

\subsection{Dijet cross sections}
\begin{figure}
\centering
\includegraphics[width=0.495\textwidth]{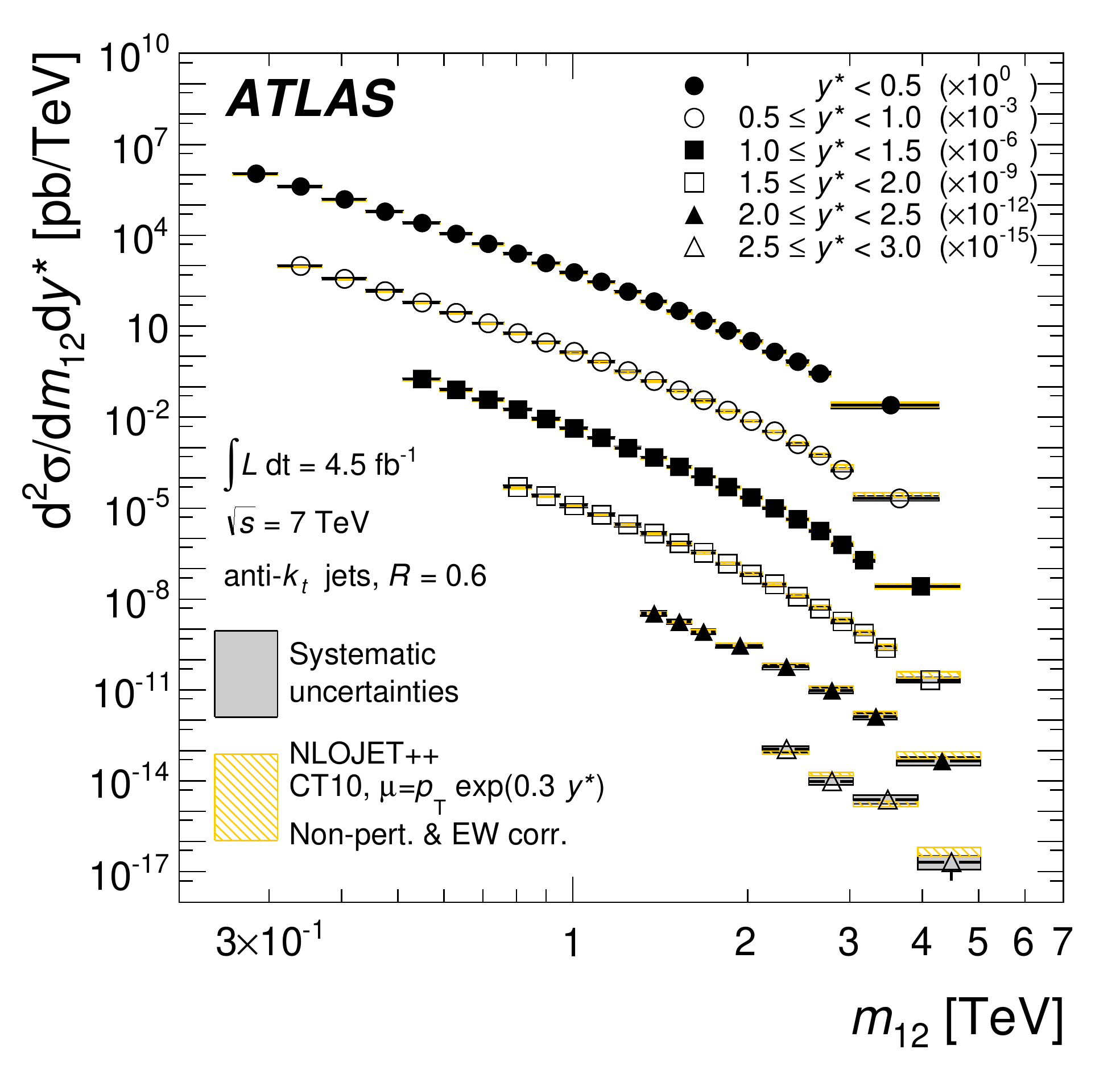}
\includegraphics[width=0.495\textwidth]{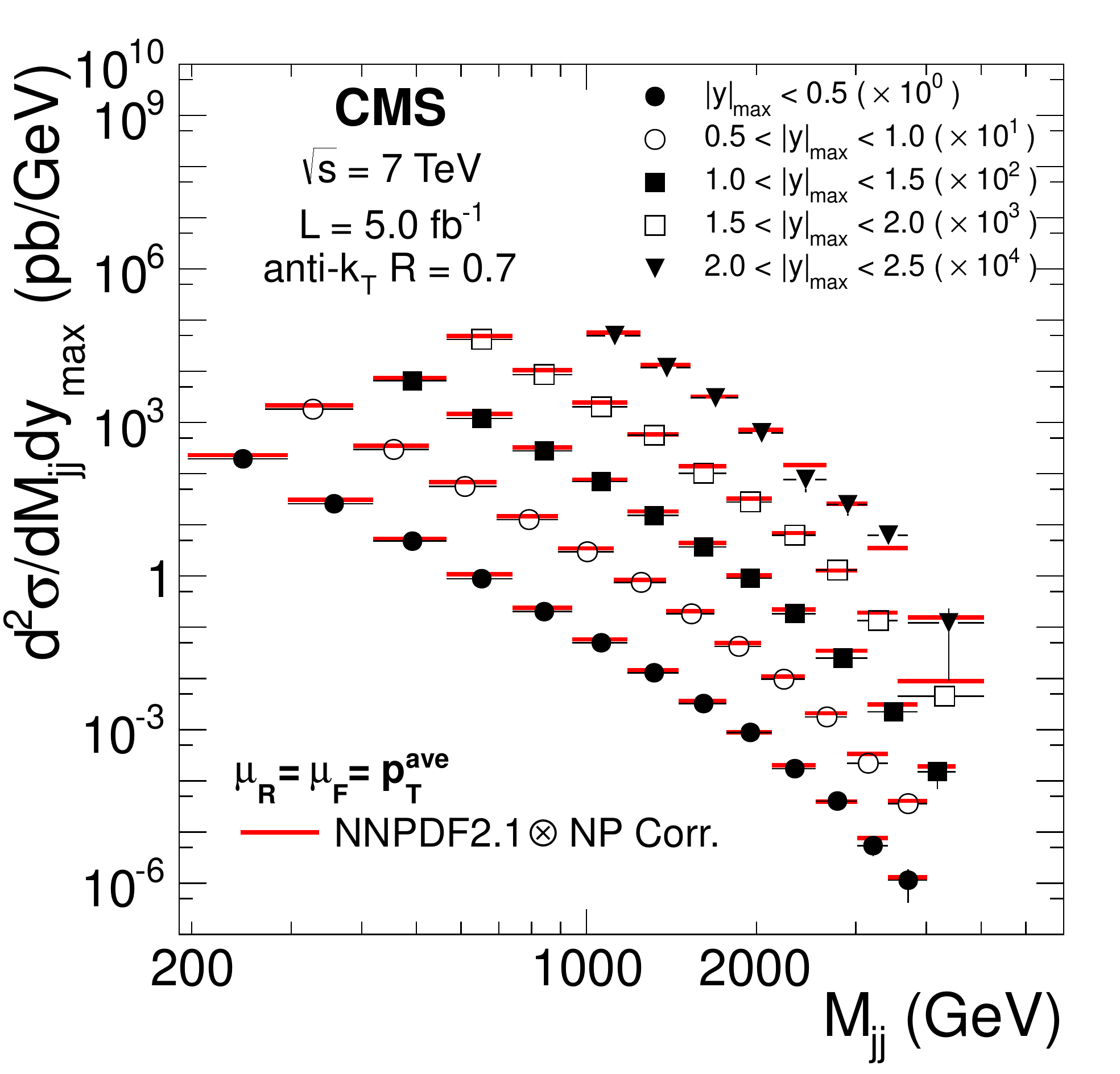}
\caption{\label{fig:CMS7TeVdijet}
Dijet double differential cross section for anti-$k_{\mathrm{t}}$ jets with values of the radius parameter $R$ = 0.6 (ATLAS) and $R$ = 0.7 (CMS) shown as a function of dijet mass in different ranges of $y*$ (ATLAS) and $y_{max}$ (CMS). To aid visibility, the cross sections are multiplied by the factors indicated in the legend.  For comparison, the NLO QCD predictions corrected for non-perturbative and electroweak effects (for ATLAS), are included. 
Taken from Ref. \citen{Aad:2013tea} and \citen{Chatrchyan:2012bja}.
}
\end{figure}

Measurements of the dijet double-differential cross sections as a function of dijet mass in various ranges of $y*$ (ATLAS) and $y_{max}$ (CMS) are shown in Fig. \ref{fig:CMS7TeVdijet} for anti-$k_{\mathrm{t}}$ jets with values of the radius parameter $R$ = 0.6 (ATLAS) and $R$ = 0.7 (CMS). The cross sections are measured up to a dijet mass of 5 TeV, and are seen to decrease quickly with increasing dijet mass. The NLO QCD calculations by \nlojetpp~which are corrected for non-perturbative and electroweak effects, are compared to the measured cross sections. No major deviation of the data from the theoretical predictions is observed over the full kinematic range, covering almost eight orders of magnitude in measured cross section values.

\begin{figure}
\centering
\includegraphics[width=0.495\textwidth]{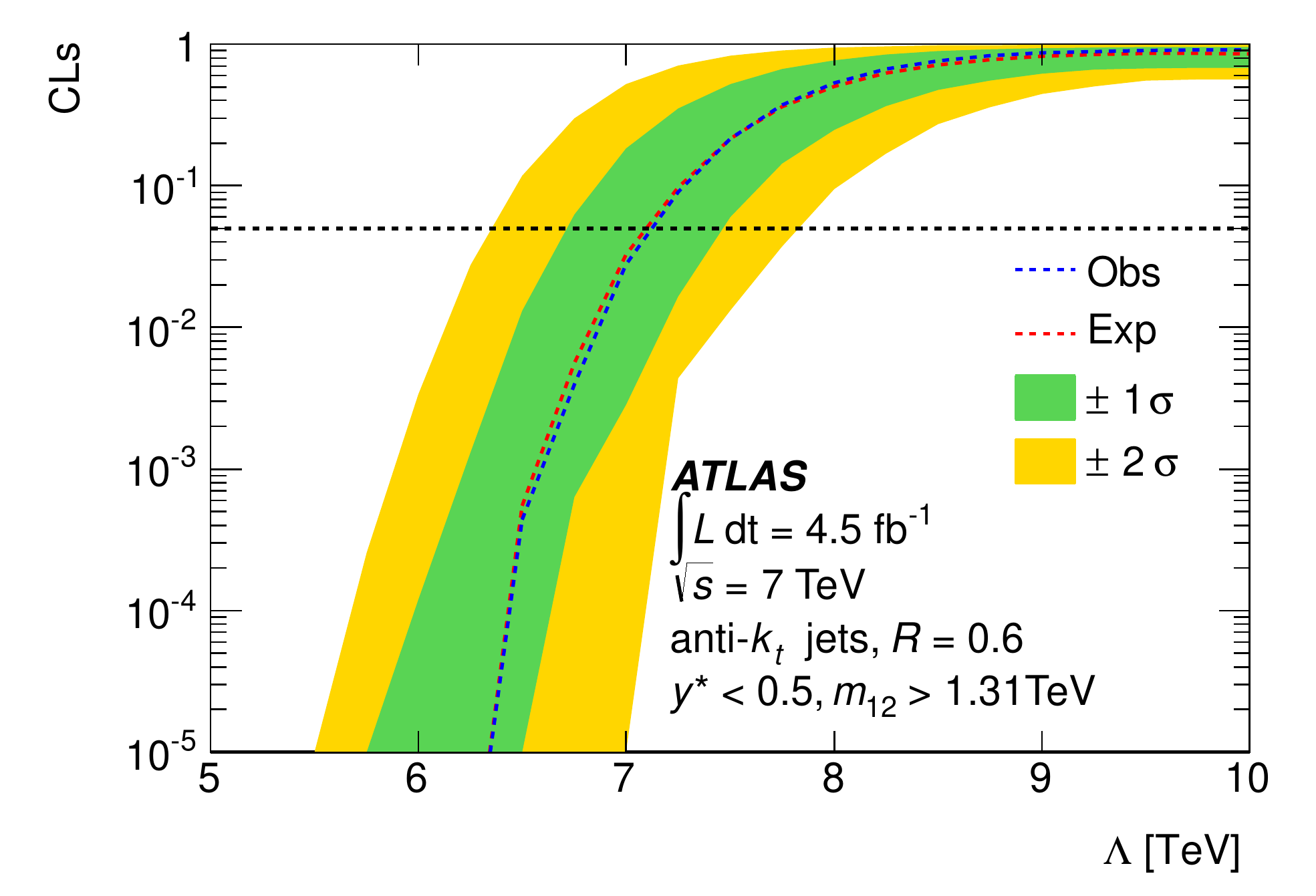}
\caption{\label{fig:ATLAS7TeVdijetLimits}
Scan of  CLs value  for NLO QCD plus contact interaction as a function of $\Lambda$.
The green (yellow) bands indicate the $\pm1\sigma(\pm2\sigma)$ regions, from pseudo-experiments of the SM background. The dashed horizontal line indicates the 95\% CL exclusion, computed using the observed (expected) p-value shown by the blue (red) dashed line. The plots correspond to the measurement with jet radius parameter $R$ = 0.6 in the range $y*< $~0.5, restricted to the high dijet-mass subsample ($m_{jj} >$ 1.31 TeV).
Taken from Ref. \citen{Aad:2013tea}
}
\end{figure}

Particularly interesting is the possibility to use the measured cross section to probe physics beyond the SM. This is done by exploring potential deviations in dijet production with respect to the SM .
In Ref. \citen{Aad:2013tea}, the model of QCD plus contact interactions is tested based on the CLs technique.
The scan of the CLs values, shown in Fig. \ref{fig:ATLAS7TeVdijetLimits}, resulted in an exclusion of the compositeness scale $\Lambda~\leq$ 7 TeV.

\newpage

\newpage

\subsection{Dijet azimuthal decorrelation}
Dijet azimuthal decorrelations for the two leading jets
in hard-scattering events can be used to study QCD radiation effects over a wide range of
jet multiplicities without the need to measure all the additional jets. Such studies are important
because an accurate description of multiple-parton radiation is still lacking in perturbative
QCD (pQCD). Experiments therefore rely on Monte Carlo event generators to take these
higher-order processes into account to get predictions for new physics and for a wide variety of precision
measurements.

\begin{figure}
\centering
\includegraphics[width=0.495\textwidth]{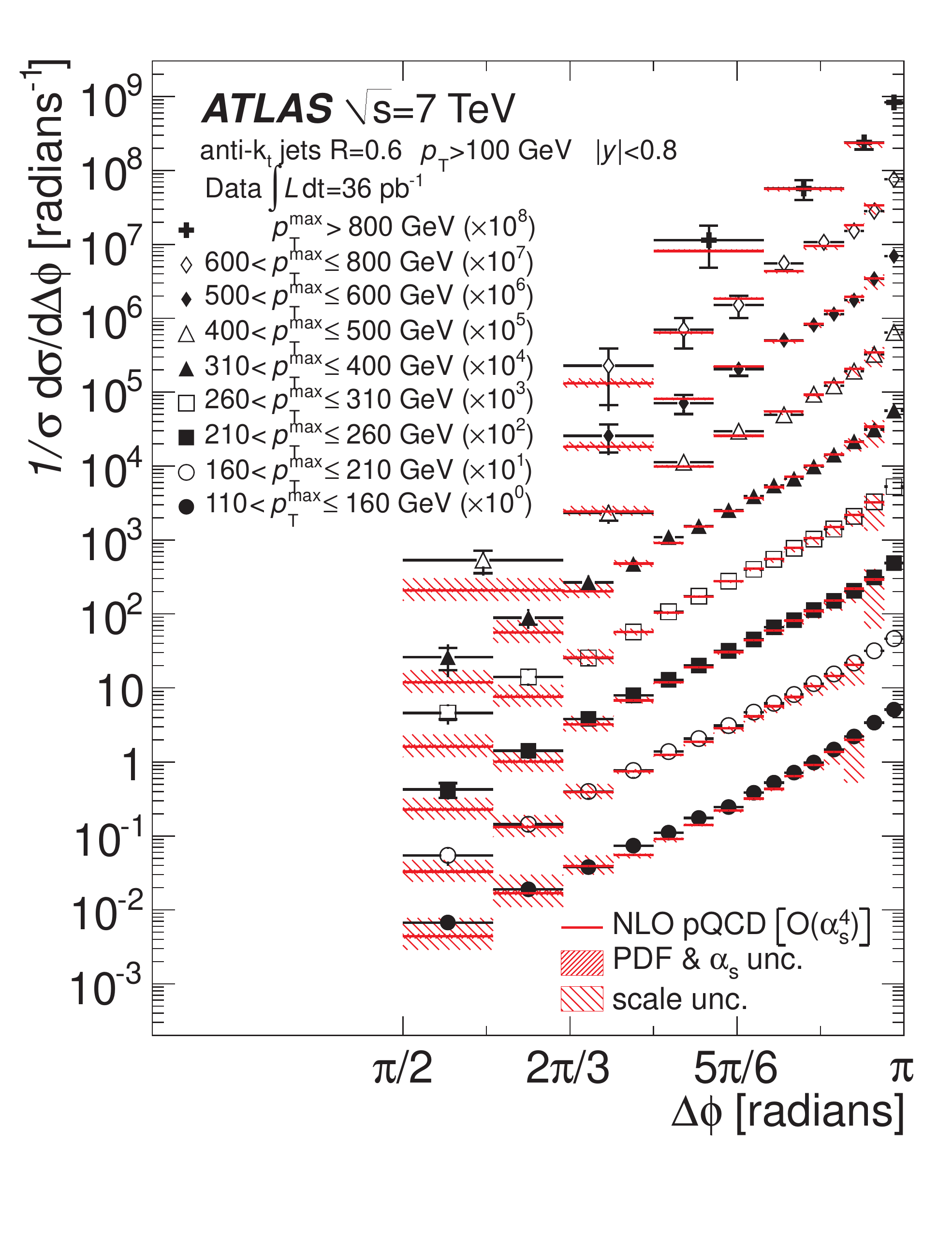}
\includegraphics[width=0.45\textwidth]{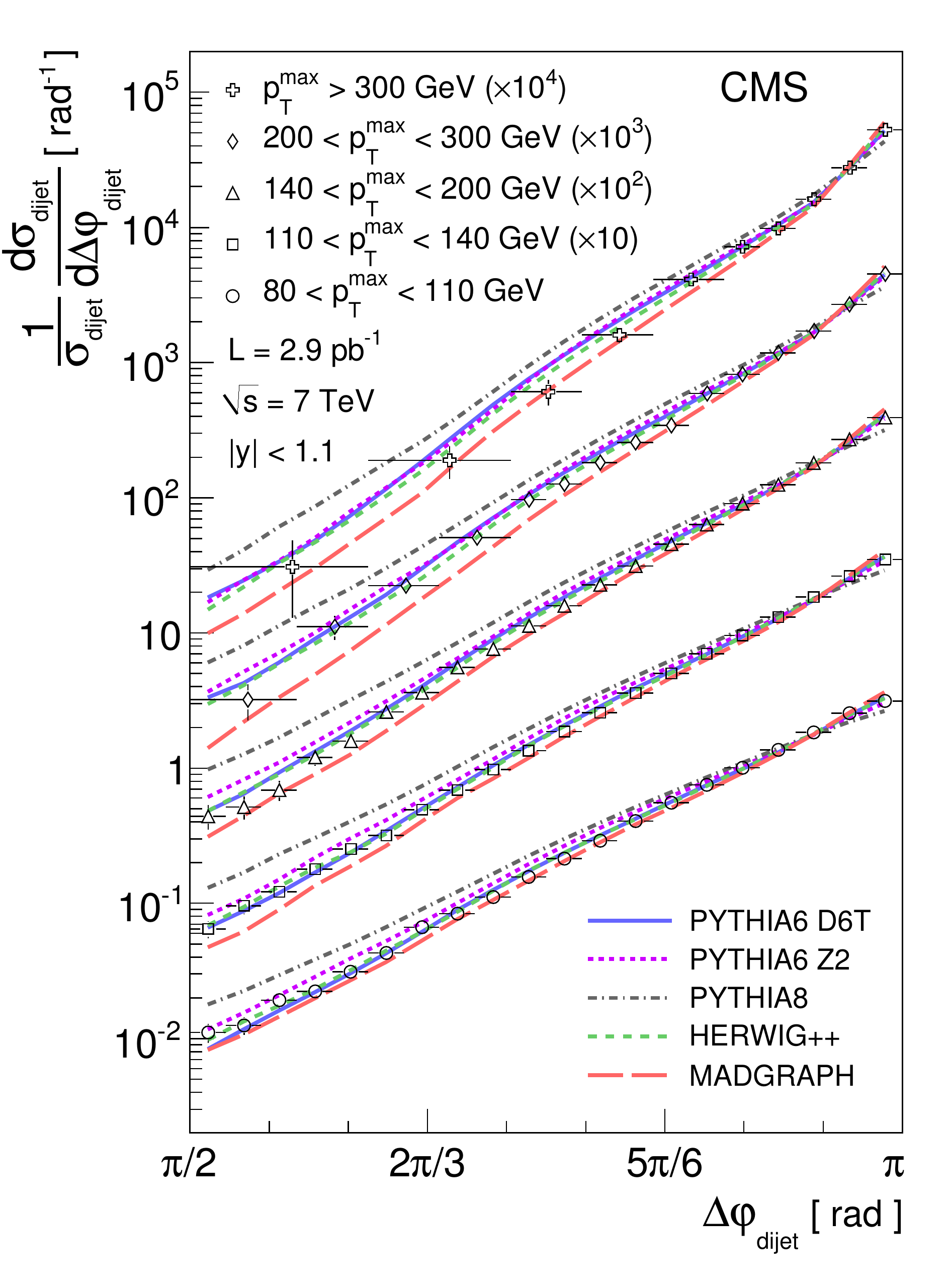}
\caption{\label{fig:DEcorr}
Dijet azimuthal decorrelation cross section  in several \pT$^{max}$ regions, scaled by the multiplicative
factors given in the figure for easier presentation. Taken from Ref. \citen{daCosta:2011ni} and \citen{Khachatryan:2011zj}.
}
\end{figure}

The differential cross sections, normalized to the integrated dijet cross section,
are shown in Fig. \ref{fig:DEcorr} for the different \pT$^{max}$ regions. The distributions are scaled by multiplicative
factors for presentation purposes.
The cross sections are strongly peaked at $\pi$ and become steeper with increasing \pT$^{max}$.

The measured cross sections are compared with a pQCD NLO calculation, corrected for non perturbative effects, and 
with MC predictions, including generators  based on leading-order matrix element multiparton
final-state predictions.  In general, the predictions are in good agreement with the measured cross section, but the pQCD NLO calculations
start to deviate from the measurement for large de-correlations ($\Delta\phi<2/3\pi$), where the effects of higher jet multiplicity, included
only partially in the pQCD calculation, play an important role. 

\clearpage
\newpage

\section{Conclusions}	
\label{sec:conclusions}
\begin{figure}
\centering
\includegraphics[width=0.6\textwidth]{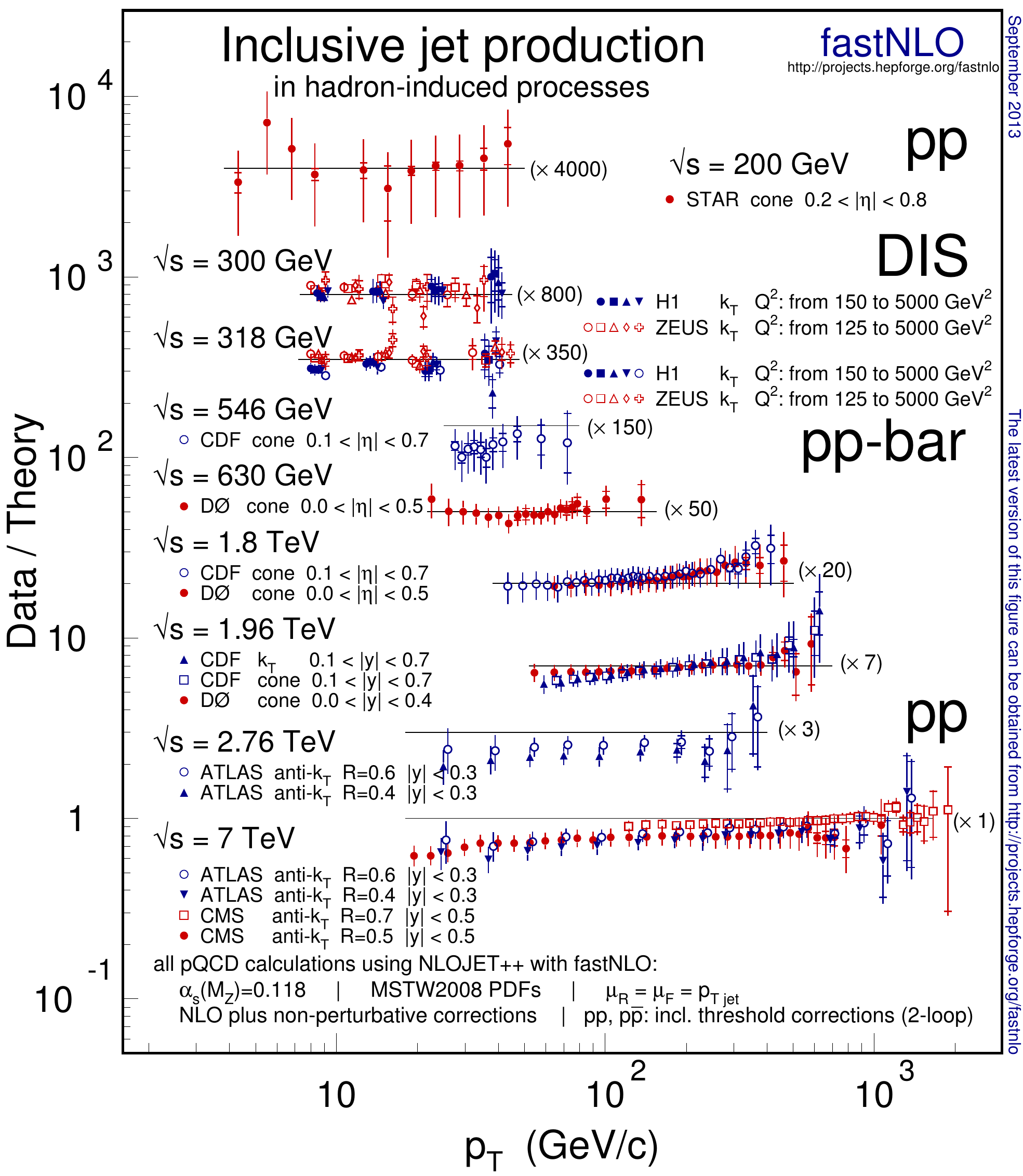}
\caption{\label{fig:jets}
A compilation of data-over-theory ratios for
inclusive jet cross sections as a function of jet transverse
momentum (\pT), measured in different hadron-induced processes
at different center-of-mass energies; from Ref. \citen{Wobisch:2011ij,Britzger:2012bs,Agashe:2014kda}. The
various ratios are scaled by arbitrary numbers (indicated
between parentheses) for better readability of the plot. The
theoretical predictions have been obtained at NLO accuracy, for
parameter choices (coupling constant, PDFs, renormalization,
and factorization scales) as indicated at the bottom of the figure.
}
\end{figure}

In the rich physics program of the experimental collaborations at the LHC, the measurements of the jet cross sections 
play an important role. These cross sections, being elegant, simple and with large rate at the LHC, are the first high-\pT~measurements 
the LHC Collaborations performed, and the first able to give information of the physics at the TeV scale. 
All the measured cross sections are in very good agreement with the Standard Model prediction. The 
improvements, both on the experimental techniques and on the theoretical prediction, allow the use of 
these measurement to study the dynamics of the interactions of quarks and gluons 
in all the relevant details. 
The LHC Run1 covered a new frontier in energy, clearly visible by comparing the jet cross section measured in
different hadron-induced processes at different center-of-mass energies shown in Fig. \ref{fig:jets}.
This new frontier will be extended in the next years, thanks to the higher beam energy reached by the Large Hadron Collider.

\newpage

\section*{Acknowledgments}
I am grateful to Gunther Dissertori for the invitation to write this review and to Bogdan Malaescu, Giovanni Marchiori, Maxime Gouzevitch and Esteban Fullana Torregrosa, for the useful feedback.
This work is partially supported by the ILP LABEX (under reference ANR-10-LABX-63 and ANR-11-IDEX-0004-02). 
\bibliographystyle{ws-ijmpa}
\bibliography{bibliography}

\end{document}